\documentclass[twocolumn]{aastex63}

\usepackage{color}
\usepackage{subfigure}
\usepackage{hyperref}

\newcommand{\ms}{\mbox{m s$^{-1}$}}
\newcommand{\cms}{\mbox{cm s$^{-1}$}}
\newcommand{\kms}{\mbox{km s$^{-1}$}}

\received{July 31, 2021}
\revised{September 30, 2021}
\accepted{October 20, 2021}

\submitjournal{AJ}

\shorttitle{EXPRES. III. $\epsilon$ Eridani}
\shortauthors{Roettenbacher et al.}

\begin{document}

\title{EXPRES. III. Revealing the Stellar Activity Radial Velocity Signature of $\epsilon$ Eridani with Photometry and Interferometry}

\correspondingauthor{R.\ M.\ Roettenbacher}
\email{rachael.roettenbacher@yale.edu}

\newcommand{\ycaa}{Yale Center for Astronomy and Astrophysics, Yale University, 46 Hillhouse Avenue, New Haven, CT 06511, USA}
\newcommand{\yale}{Department of Astronomy, Yale University, 52 Hillhouse Avenue, New Haven, CT 06511, USA}
\newcommand{\sfsu}{Department of Physics and Astronomy, San Francisco State University, 1600 Holloway Avenue, San Francisco, CA 94132, USA}
\newcommand{\lowell}{Lowell Observatory, 1400 W. Mars Hill Road, Flagstaff, AZ 86001, USA}
\newcommand{\yalephysics}{Department of Physics, Yale University, 217 Prospect St, New Haven, CT 06511, USA}
\newcommand{\eso}{European Southern Observatory, Alonso de C\'ordova 3107, Vitacura, Casilla 19001, Santiago, Chile}
\newcommand{\tsu}{Tennessee State University, Center of Excellence in Information Systems, Nashville, TN 37203, USA}
\newcommand{\michigan}{Department of Astronomy, University of Michigan, Ann Arbor, MI 48109, USA}
\newcommand{\exeter}{Astrophysics Group, Department of Physics \& Astronomy, University of Exeter, Stocker Road, Exeter EX4 4QL, UK}
\newcommand{\grenoble}{Institut de Planetologie et d’Astrophysique de Grenoble, Grenoble F-38058, France}
\newcommand{\ua}{Steward Observatory, Department of Astronomy, University of Arizona, 933 N. Cherry Avenue, Tucson, AZ, 85721, USA}
\newcommand{\chara}{The CHARA Array of Georgia State University, Mount Wilson Observatory, Mount Wilson, CA 91203, USA}
\newcommand{\leuven}{Institute of Astronomy, KU Leuven, Celestijnenlaan 200D, B-3001, Leuven, Belgium}
\newcommand{\owu}{Department of Physics and Astronomy, Ohio Wesleyan University, Delaware, OH 43015, USA}
\newcommand{\toledo}{Department of Physics and Astronomy, University of Toledo, Toledo, OH 43606}
\newcommand{\nau}{Department of Astronomy and Planetary Science, Northern Arizona University, Flagstaff, AZ 86011}
\newcommand{\uw}{Department of Astronomy, University of Washington, Seattle, WA, 98195}
\newcommand{\bu}{Institute for Astrophysical Research,
Boston University, 725 Commonwealth Ave.,
Boston, MA 02215}
\newcommand{\aao}{Australian Astronomical Optics, Faculty of Science and Engineering, Macquarie University, Macquarie Park, NSW 2113, Australia}
\newcommand{\umd}{Department of Astronomy, University of Maryland,
College Park, MD 20742}

\author[0000-0002-9288-3482]{Rachael M.\ Roettenbacher}
\affiliation{\ycaa}
\affiliation{\yale}

\author[0000-0001-9749-6150]{Samuel H.\ C.\ Cabot}
\affiliation{\yale}

\author[0000-0003-2221-0861]{Debra A.\ Fischer}
\affiliation{\yale}

\author[0000-0002-3380-3307]{John D.\ Monnier}
\affiliation{\michigan}

\author[0000-0003-4155-8513]{Gregory W.\ Henry}
\affiliation{\tsu}

\author{Robert O.\ Harmon}
\affiliation{\owu}

\author[0000-0003-0529-1161]{Heidi Korhonen}
\affiliation{\eso}

\author[0000-0002-9873-1471]{John M.\ Brewer}
\affiliation{\sfsu}

\author[0000-0003-4450-0368]{Joe Llama}
\affiliation{\lowell}

\author[0000-0003-2168-0191]{Ryan R.\ Petersburg}
\affiliation{\yalephysics}

\author[0000-0002-3852-3590]{Lily Zhao}
\affiliation{\yale}

\author[0000-0001-6017-8773]{Stefan Kraus}
\affiliation{\exeter}

\author[0000-0002-0493-4674]{Jean-Baptiste Le Bouquin}
\affiliation{\grenoble}

\author[0000-0002-2208-6541]{Narsireddy Anugu}
\affiliation{\ua}

\author[0000-0001-9764-2357]{Claire L.\ Davies}
\affiliation{\exeter}

\author[0000-0002-3003-3183]{Tyler Gardner}
\affiliation{\michigan}

\author[0000-0001-9745-5834]{Cyprien Lanthermann}
\affiliation{\leuven}
\affiliation{\chara}

\author[0000-0001-5415-9189]{Gail Schaefer}
\affiliation{\chara}

\author[0000-0001-5980-0246]{Benjamin Setterholm}
\affiliation{\michigan}

\author[0000-0002-2361-5812]{Catherine A.\ Clark}
\affiliation{\nau}

\author[0000-0001-6158-1708]{Svetlana G.\ Jorstad}
\affiliation{\bu}

\author[0000-0003-0120-0808]{Kyler Kuehn}
\affiliation{\lowell}
\affiliation{\aao}

\author{Stephen Levine}
\affiliation{\lowell}

\begin{abstract}

The distortions of absorption line profiles caused by photospheric brightness variations on the surfaces of cool, main-sequence stars can mimic or overwhelm radial velocity (RV) shifts due to the presence of exoplanets. The latest generation of precision RV spectrographs aims to detect velocity amplitudes $\lesssim 10$ \cms, but requires mitigation of stellar signals.  Statistical techniques are being developed to differentiate between Keplerian and activity-related velocity perturbations.  Two important challenges, however, are the interpretability of the stellar activity component as RV models become more sophisticated, and ensuring the lowest-amplitude Keplerian signatures are not inadvertently accounted for in flexible models of stellar activity.  For the K2V exoplanet host $\epsilon$ Eridani, we separately use ground-based photometry to constrain Gaussian processes for modeling RVs and TESS photometry with a light-curve inversion algorithm  to reconstruct the stellar surface. From the reconstructions of TESS photometry, we produce an activity model, which reduces the rms scatter in RVs obtained with EXPRES from 4.72 \ms\ to 1.98 \ms. We present a pilot study using the CHARA Array and MIRC-X beam combiner to directly image the starspots seen in the TESS photometry.  With the limited phase coverage, our spot detections are marginal with current data but a future dedicated observing campaign should allow for imaging, as well as the stellar inclination and orientation with respect to its debris disk to be definitely determined.  This work shows that stellar surface maps obtained with high cadence, time-series photometric and interferometric data can provide the constraints needed to accurately reduce RV scatter.

\end{abstract}


\section{Introduction} 
\label{sec:intro}

Radial velocity surveys for exoplanets have not yet been able to detect planets with similar masses and radii to those of the Earth in Earth-like orbits around Sun-like stars.  However, the latest generation of spectrographs are designed to reduce instrumental error sources with the goal of isolating the stellar signals that obstruct the detection of low-amplitude velocity signals.  With extreme precision radial velocity (EPRV) surveys of solar analogs, features such as starspots on the stellar surface produce temporal variations in the shapes of line profiles that add time-correlated variations to the center of mass radial velocity (RV) measurements and must be properly accounted for.   

A number of phenomena contribute to the absorption line profile signatures caused by stellar activity that result in RV shifts.  The convective envelope of cool, Sun-like stars is composed of cells, or granules, in which hot stellar material rises and then falls as it cools. Localized, strong magnetic fields suppress convection and manifest as bright regions, including faculae, plages, and networks, and dark starspots on the stellar surface.  As these features rotate in and out of view, they create rotationally-modulated absorption line signatures that lead to periodic RV signatures.  Faculae, which contribute RV amplitudes $< 1$ \ms{}, are bright with respect to the photosphere and are most apparent when near the limb of the stellar disk \citep{hay16}. Plages and networks are also bright with respect to the photosphere, but have a more significant RV amplitude contribution on the order of a few \ms, as seen on the Sun \citep{mil19}. Starspots are dark features against the photosphere and can cause line profile distortions that contribute a wide range of RV amplitudes---from less than 1 \ms\ to several \kms\ for large starspots \citep{roe15b,hay16}.  Even for chromospherically quiet stars, these signatures can overwhelm the  $\sim 10$ \cms\ RV signature of an Earth analog in the habitable zone.

Ongoing efforts to characterize and isolate stellar activity signals include: modeling stellar activity with flexible correlated noise models \citep[e.g., Gaussian processes;][]{hay14,gil20}; statistically identifying stellar activity \citep[e.g., PCA or F-statistic;][]{davis2017, hol21}; estimating RV variations from a spot model applied to photometry \citep{aig12, dum14}; extracting signatures of stellar activity from the cross-correlation function used to measure RVs \citep{col21}; and using Doppler imaging of young stars to filter stellar activity from line profiles \citep{hei21}.  Several additional or related methods are compared by \citet{dum2017} and Zhao et al.\ (2021, in preparation).  These methods are based upon photometric and/or spectroscopic data and aim to account for RV scatter due to photospheric activity. 

There are a number of ways to reconstruct the stellar surfaces to resolve some of the surface structure contributing to RV signatures.  Photometric light-curve inversion uses one or more light curves of a star showing rotational variability to reconstruct the stellar surface \citep{sav08,har00,luo19}.  This method makes no assumptions on the spot size, shape, or number, and has been shown to be reliable when compared to simulations and interferometric images \citep{har00,roe17b}. While the method gives starspot longitudes, degeneracies only allow for the determination of relative starspot latitudes when multiple features are present, and  degeneracies will remain between the size of a starspot and its absolute latitude \citep[e.g.,][]{har00}. Doppler imaging is a method to reconstruct relatively large surface spots using high-resolution spectra and can better determine starspot latitude \citep[e.g.,][]{vog87,ric89}, but a degeneracy between the hemispheres remains.  Doppler imaging requires a rotational velocity of $\sim10$~\kms, restricting the stars to which the method can be applied \citep[example applications of the method include][]{kor21,sen21}. Interferometric aperture synthesis imaging is a third method that can be used to reconstruct the stellar surface.  Because interferometric imaging allows for stars and their spots to be imaged as they appear on the sky, the degeneracies in  starspot latitudes of other techniques are resolved.  With  observations spanning a stellar rotation period, the stellar inclination and the star's orientation on the sky---the position angle of the axis of rotation---can be measured. Resolving the surfaces of stars is currently only possible for bright stars that have angular diameters  $\sim2$~mas or more and relatively large starspots \citep[e.g.,][]{roe16a,roe17b,par21,mar21}. 

Knowing the brightness inhomogeneities, such as starspots, present at the time of RV observations can provide a way to separate their impact on line profiles and thus RV signatures from those of the planets. This has provided motivation for solar telescopes that measure radial velocities for disk-integrated spectra of the Sun \citep{col21, dum2021}. Here, we use a similar approach to analyze the star $\epsilon$ Eridani (HD 22049, TIC 118572803).  The closest K2 dwarf to the Sun, $\epsilon$ Eri is  at a distance of $3.220 \pm 0.004$~pc \citep{gai21}.  It is a bright \citep[$V = 3.73$, $H = 1.75$;][]{duc02}, main-sequence star with a radius of $0.74 \pm 0.01~R_\odot$ derived from interferometry \citep{dif07,bai12}.  It is known to be active with variable starspots and detected global activity cycles \citep{met13}.  The star has a rotation period of approximately 11 days, detected with a variety of techniques, including MOST photometry and radial velocity variations \citep{gig16}, modulation of Ca II H\&K  measurements \citep{hem16}, and photometry \citep{lan14}. $\epsilon$ Eri is also a known exoplanet host star \citep{hat00, maw19} with an RV-detected $0.8~M_\mathrm{Jup}$ planet in a $7.4$-year orbit.  

Here, we utilize stellar surface images to isolate the associated RV signatures in EPRV data.  In Section \ref{observations}, we present the spectroscopic, interferometric, and photometric observations used in this work. Spectroscopic observations include an extensive, multi-decade baseline archival data set, as well as new, high-precision RV measurements. In Section \ref{sec:gp}, we detail our Gaussian process analysis, infer attributes of the stellar activity, and confirm orbital parameters of the known planetary companion. In Section \ref{sec:images}, we reconstruct the stellar surface with a light-curve inversion algorithm and model interferometric data obtained on two nights. In Section~\ref{sec:specsims}, the  surface reconstructed from a light curve is combined with a disk model that simulates the stellar spectrum. We discuss the simulated stellar spectra and show how it successfully accounts for a significant portion of scatter in contemporaneous RV measurements.  The model's efficacy, additional considerations, and future work are discussed in Section~\ref{sec:disc}.

\section{Observations}
\label{observations}

$\epsilon$ Eri has been extensively observed owing to its brightness and proximity to the Sun.  In this section, we describe the  observations that were used for this work, including brief descriptions of the archival data. 

\subsection{EXPRES Spectroscopy}
\label{expres}

High-resolution spectra of $\epsilon$ Eri were acquired with the Extreme PREcision Spectrometer \citep[EXPRES;][]{jur16} commissioned at the 4.3-m {\it Lowell Discovery Telescope} \citep[LDT;][]{Levine2012}. EXPRES is an optical spectrograph optimized for wavelengths $380-780$ nm, reaching a typical resolving power of $R\sim137,500$. Additional specifications can be found in the works detailing the radial velocity pipeline, instrument performance verification, and first science results \citep{pet20, bla20, bre20}. EXPRES attains $\sim 30$ cm s$^{-1}$ radial velocity precision for spectra of slowly-rotating, main-sequence FGK-stars when the signal-to-noise ratio reaches $250$ at 550~nm. While the lowest root-mean-square (rms) RV scatter is observed around chromospherically inactive stars, we observed $\epsilon$ Eri as an interesting case-study for characterizing and mitigating RV jitter in a moderately active star.

We obtained 164 RVs of $\epsilon$ Eri on 39 distinct nights between 2019 August 15 and 2020 November 13 (see Table \ref{tab:newrvs}). Besides one seasonal gap, $\epsilon$ Eri was observed every $3-10$ nights. Stellar activity dominates the RVs, yielding an rms scatter of 6.6 m s$^{-1}$ with an average single measurement uncertainty of $35$ cm s$^{-1}$. 

\begin{deluxetable}{l c c}
\tabletypesize{\scriptsize}
\tablecaption{EXPRES Radial Velocities}
\tablewidth{0pt}
\tablehead{
\colhead{Reduced Julian Date} &  \colhead{$v$} & \colhead{$\sigma_v$}  \\  \colhead{(RJD = JD - 2400000.0) } &  \colhead{ (\ms) } & \colhead{(\ms)}
}
\startdata
 58710.983924   &   -9.762  &  0.482 	\\
 58710.985348   &  -10.127  &  0.466	\\
 58710.986526   &   -9.149  &  0.692	\\
 58716.996130   &  -14.496  &  0.387 	\\
 58716.997549   &  -10.840  &  0.380	\\
$\cdots$    &   $\cdots$    &   $\cdots$    \\
\enddata
  \tablecomments{This table is available in machine-readable form.}
\label{tab:newrvs}
\end{deluxetable}

\subsection{Archival Spectroscopy}
\label{sec:archivalspectra}

\citet{maw19} conducted an extensive analysis of archival $\epsilon$ Eri RVs spanning 30 years, coupled with direct imaging of the system. They placed tight constraints on the orbit of the 7-year planet first identified by \citet{hat00}. The study made use of over 450 RVs obtained with the following instruments (their respective observatories are listed immediately after them): High Resolution Echelle Spectrometer (HIRES)/Keck, Levy Automated Planet Finder (APF)/Lick,
Hamilton/Lick, Coudé Echelle/La Silla, and High Accuracy Radial Velocity Planet Searcher (HARPS)/La Silla. We refer the reader to \citet{maw19} and \citet{zec13} for details on these RVs. Our analysis of the archival data additionally includes CHIRON RVs acquired in 2014 \citep{gig16}.

\subsection{TESS Photometry}
\label{sec:TESS}

The Transiting Exoplanet Survey Satellite \citep[TESS;][]{ric14} observed $\epsilon$ Eri during Sector 31 (2020 October 21 through 2020 November 19). The 2-minute cadence simple aperture photometry (SAP) light curves were obtained through the Barbara A.\ Mikulski Archive for Space Telescopes (MAST).  

To account for scattered light and other systematic issues, we typically remove cotrending basis vectors (CBVs) from the SAP light curves \citep[following the procedure used in][]{roe18,cab21}.  However, we did not remove any CBVs from the Sector 31 light curve, as the CBVs provided through the MAST archive for Sector 31 (with a file creation date of 2020 December 12) add trends to the data or increase the noise.  The shape of the SAP light curve is consistent with the nearly contemporaneous, ground-based light curve described in Section \ref{sec:APT}, suggesting stellar variability is likely the dominant signature.   We removed data flagged as bad by the TESS pipeline (a non-zero quality flag).  There were no usable data provided for the SAP light curve between JD 2459155.8948312--2459158.8670810 due to scattered light from the Moon and the transition between orbits of the satellite.  $\epsilon$ Eri was observed with Camera 1, which was strongly impacted by scattered light at the time of these removed data points according to the Sector 31 Data Release Notes (DR47)\footnote{\url{https://archive.stsci.edu/missions/tess/doc/tess_drn/tess_sector_31_drn47_v02.pdf}}.  We removed a total of 2376 observations from the original 18314 observations.   

\subsection{Automated Photoelectric Telescope Photometry}
\label{sec:APT}

$\epsilon$ Eri was observed with two Automatic Photoelectric Telescopes (APTs) at Fairborn Observatory \citep{hen99}.  600 observations were obtained with the T8 (0.8~m) telescope from 2013 October 11 through  2020 February 20 and 64 observations on the T4 (0.75~m) telescope from 2020 October 30 through 2020 November 30.  
The differential photometric observations were obtained in Str\"omgren $b$ and $y$ bandpasses, but are presented as a combined $(b + y)/2$ bandpass. 
The stars HD 22243 and HD 23281 were used as comparison stars to ensure the variable signature was that of $\epsilon$ Eri.  

We removed the long-term trends in both of these data sets.  To do so, we smoothed the data with a Gaussian kernel with a full width at half maximum of 50 days and removed that signature to leave only the signature of the starspots that were rotating in and out of view.   These data are included in Tables \ref{tab:APT_T8} and \ref{tab:APT_T4}. 

\begin{deluxetable}{l c c}
\tabletypesize{\scriptsize}
\tablecaption{Str\"omgren $(b+y)/2$ differential photometric data of $\epsilon$~Eri with the APT T8}
\tablewidth{0pt}
\tablehead{
\colhead{Modified Julian Date} &  \colhead{$(b+y)/2$ Differential} & \colhead{Trend Removed}  \\  \colhead{(MJD = JD - 2400000.5) } &  \colhead{Magnitude } & \colhead{}
}
\startdata
56576.813	&	-1.98170	&	-1.97666	\\
56576.901	&	-1.98095	&	-1.97666	\\
56577.815	&	-1.98250	&	-1.97661	\\
56584.791	&	-1.97605	&	-1.97630	\\
56584.875	&	-1.97405	&	-1.97629	\\
$\cdots$    &   $\cdots$    &   $\cdots$    \\
\enddata
  \tablecomments{This table is available in machine-readable form.}
\label{tab:APT_T8}
\end{deluxetable}

\begin{deluxetable}{l c c}
\vspace{-1cm}
\tabletypesize{\scriptsize}
\tablecaption{Str\"omgren $(b+y)/2$ differential photometric data of $\epsilon$~Eri with the APT T4}
\tablewidth{0pt}
\tablehead{
\colhead{Modified Julian Date} &  \colhead{$(b+y)/2$ Differential} & \colhead{Trend Removed} \\ \colhead{(MJD = JD - 2400000.5) } &  \colhead{Magnitude } & \colhead{}
}
\startdata
59152.758	&	-1.96490	&	-1.96735	\\
59152.820	&	-1.96360	&	-1.96735	\\
59152.844	&	-1.96520	&	-1.96735	\\
59152.878	&	-1.96450	&	-1.96735	\\
59152.941	&	-1.96405	&	-1.96735	\\
$\cdots$    &   $\cdots$    &   $\cdots$    \\
\enddata
  \tablecomments{This table is available in machine-readable form.}
\label{tab:APT_T4}
\end{deluxetable}

\subsection{MIRC-X Interferometry}
\label{sec:mircx}

Two sets of interferometric observations were obtained on 2020 November 2 (JD 2459155.77) and 2020 November 5 (JD 2459158.87) at the Center for High Angular Resolution Astronomy (CHARA) Array \citep{ten05} using the Michigan InfraRed Combiner-eXeter \citep[MIRC-X;][]{anu20}.  The CHARA Array consists of six 1-m telescopes in a Y-shaped array with non-redundant baselines ($B$) extending from 34 to 330 m \citep{ten05}. The snapshot observations were obtained in $H$-band (spanning $1.5-1.8 \mu$m; $\lambda/2B \approx 0.5$~mas) in the grism mode ($R\sim190$).  The observations of $\epsilon$ Eri were followed by those of the calibration star, HD~26912 \citep[uniform disk angular diameter measured in $H$-band with no limb darkening, $\theta_{\mathrm{UD},H} = 0.285 \pm 0.026$~mas;][]{che16}.

We reduced the interferometric observations with the standard MIRC-X reduction pipeline (version 1.3.5)\footnote{\url{https://gitlab.chara.gsu.edu/lebouquj/mircx_pipeline}} with the default reduction parameters, and we set the number of coherent coadds (ncoh) to 10, the flux threshold to 5, and the signal-to-noise threshold to 3.  We calibrated the data with a modified version of the calibration software for the previous Michigan InfraRed Combiner \citep[MIRC;][]{mon12}, which allowed for the removal of bad data that the automated pipeline would not properly flag.  

\section{Gaussian Process Analysis} \label{sec:gp}

Gaussian Processes (GPs) have become a frequently-used tool in analyses of RV time-series since case studies by \citet{hay14} and \citet{raj15}. Specifically, they are often used as a flexible model of correlated structure attributed to stellar activity. GPs are advantageous because of their analytically tractable likelihood function, and simple parametrization through their covariance function \citep[further details surrounding GPs may be found in][]{ram2006}. 

We recently applied GPs to high-precision EXPRES RVs of the bright, Sun-like star HD 101501 and demonstrated that a combination of high-amplitude stellar activity and sparse observing cadence inhibit the detection of low-mass planets \citep{cab21}. We showed that high-cadence RVs drastically improved the detectable parameter space for planets around active stars, and that simultaneous photometry provides important constraints on stellar activity that are otherwise difficult to infer from RVs alone. We used the same framework in the following analysis. We opted to use the \texttt{celerite} quasi-periodic covariance kernel \citep{fm2017} with hyperparameters $\{B, C, L, P_{\rm GP}\}$, corresponding to amplitude of the covariance, weighting of the sinusoidal term, decay parameter, and recurrence timescale, respectively. The covariance matrix is constructed for pairs of timestamps ($t_i$, $t_j$) and has the form
\begin{equation}
    K_{ij} = \frac{B}{2+C}e^{-|t_i-t_j|/L}\Big[\cos{\frac{2\pi|t_i-t_j|}{P_{\rm GP}}} + (1 + C)\Big],
    \label{eqn:k2}
\end{equation}
where all hyperparameters are positive \citep{fm2017}. \texttt{celerite} covariance matrices may be inverted with reduced complexity, which makes its GPs appropriate for the extensive RV data set considered here.  The kernel is markedly faster than the commonly-used quasi-periodic kernel \citep[e.g.,][]{hay14}, which may be implemented with the \texttt{george} package \citep{amb2015}. In contrast to the \texttt{george} quasi-periodic kernel, the \texttt{celerite} quasi-periodic kernel is not mean-square differentiable \citep{ram2006}, and its covariance decreases faster for a fixed decay timescale parameter (denoted $L$ for the \texttt{celerite} kernel, and $\lambda_e$ for the \texttt{george} kernel.)  Our fitting process makes use of the nested sampler \texttt{PyMultinest} \citep{fer2008, fer2009, buc2014, fer2019}.  Sampling parameters include 4000 live points (except for our GP \& 2-Planet model, which used 6000 live points), a sampling efficiency of 0.6, and evidence tolerance of 0.5. For each model, the sampler was run three times in order to obtain a median and standard deviation on the log-evidence $\ln\mathcal{Z}$. We also confirmed that the inferred parameters were consistent across the three runs (within $1\sigma$ uncertainties).

We note that \citet{maw19} exclude a GP from their model after showing it is statistically disfavored by the Bayesian Information Criterion (BIC), and that it does not significantly impact derived parameters. Our study focuses on characterizing stellar activity, as well as searching for Keplerian signals that have amplitude less than that of the stellar activity signal. In particular, we are interested in the effect of rotationally modulated signals from starspots, which can be spatially identified with a light-curve inversion algorithm (see Section \ref{sec:li}). Therefore, we retain a quasi-periodic GP in our model.

\subsection{Photometry Preconditioning}

\begin{figure} 
\vspace{-0.6cm}
\hspace{-0.7cm}
\includegraphics[scale = 0.35]{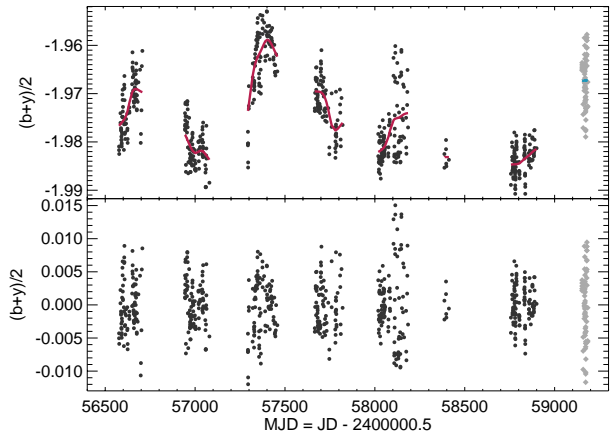}
\vspace{-1cm}
\caption{Ground-based APT light curve of $\epsilon$ Eri. Top:  Differential $(b+y)/2$ photometry (in magnitudes) from the T8 (black circles) and the T4 (gray diamonds, latest data set) telescopes.  The long-term trends are shown in red for the T8 and blue for the T4 data. Bottom:  The same photometry as above, but with the long-term trends removed.  The signature that remains is assumed to be  rotational variation and not overall brightening or dimming trends.  }
\label{fig:APT_LC}
\end{figure}

The eight-year data set of ground-based APT photometry, described in Section \ref{sec:APT} and shown in Figure \ref{fig:APT_LC}, was used to obtain posterior distributions on GP hyperparameters. Broad, log-uniform ($\mathcal{LU}$) or uniform ($\mathcal{U}$) priors were assigned to the GP hyperparameters $B, L, P_{\rm GP}$, and $C$. The fit involved a jitter term $s$ added in quadrature to all uncertainties, as well as a global offset $\gamma$. The jitter and offset terms were given a broad $\mathcal{LU}$ and broad $\mathcal{U}$ prior, respectively. TESS photometry did not span enough rotations to constrain the GP parameters, and was not used in this analysis. Posterior samples are shown in Appendix \ref{appendix:posterior}, along with distribution medians and $16\%$ to $84\%$ confidence intervals. The periodic timescale $P_{\rm GP}$ is very well constrained at $11.4\pm0.2$~days. The decay timescale $L$ is found to be $40^{+20}_{-10}$ days. The constraints on these hyperparameters are based on the physical process of spots and faculae evolving on the rotating stellar surface. The parameters $P_{\rm GP}$ and $L$ approximately correspond to the stellar rotation period at the typical latitude of spots and the typical spot lifetime, respectively. While not imposed explicitly as a prior, the results indicate the decay timescale is longer than the periodic timescale.

There is no accurate, analytical model for the influence of stellar activity on radial velocity measurements; however, RV studies searching for exoplanets have successfully used GPs regressed to the RV measurements \citep{hay14} to model the activity component. Since RVs are usually more sparse than photometry, it is useful to first regress a GP to photometry and determine the best-fit hyperparameters (as we have done above), and then regress another GP to the RV measurements where most of the hyperparameters are fixed to the photometry-based best-fit values \citep{hay14}. We adopted the same strategy as \citet{hay14} in our RV analysis by fixing three GP hyperparameters to their {\it maximum a posteriori} (MAP) values from the photometry fit: $\log C = -6.7$, $L = 36.5$ days, and $P_{\rm GP} = 11.4$ days. These fixed hyperparameters are listed in Table~\ref{tab:rvresults} for the ``preconditioned'' models. The amplitude hyperparameter $B$ was left free, since the photometry and RVs naturally concern different units. \citet{kos2020} demonstrated that best-fit hyperparameters for quasi-periodic GPs were similar when regressed separately to contemporaneous RVs and photometry of the Sun. While this result supports the notion of preconditioning with photometry, a variety of factors such as spot distributions, instrument systematics, or non-contemporaneous time series could plausibly result in different best-fit hyperparameters. To check the efficacy of preconditioning GP hyperparameters on the photometry fit, we explored a GP \& 1-Planet model that left all GP hyperparameters as free (Table~\ref{tab:rvresults}, first column). The orbital parameters were largely unaffected. Interestingly, the inferred periodic timescale $P_{\rm GP} = 11.8\pm0.4$ days is very close to the photometry preconditioning best-fit value ($P_{\rm GP} = 11.4$ days), demonstrating a strong periodic component in the RV data set arising from stellar rotation, and that the constraint proposed by \citet{kos2020} is unnecessary in our case. The decay timescale $L = 21^{+9}_{-5}$ days is roughly half that of the best-fit GP photometry value ($L = 36.5$ days), allowing the GP to more rapidly evolve between subsequent stellar rotations; however, the two estimates overlap at their 1$\sigma$ upper and lower limits, respectively.

\begin{deluxetable*}{l c r r r r r r}
\tabletypesize{\scriptsize}
\tablecaption{GP Analysis Results}
\tablewidth{0pt}
\tablehead{
\colhead{Parameter} & \colhead{Units} & \colhead{GP \& 1-Planet} & \colhead{} & \colhead{GP \& 1-Planet} & \colhead{} & \colhead{GP \& 2-Planets} & \colhead{} \\
\colhead{}  & \colhead{} & \colhead{(free)} & \colhead{} & \colhead{(preconditioned)} & \colhead{} & \colhead{(preconditioned)} & \colhead{} 
}
\startdata
  $B$                & (\ms)$^2$  &  $59^{+17}_{-8}$             &  (54)      & $69^{+11}_{-8}$            &  (66)      & $70^{+17}_{-8}$            &  (60)         \\
  $\ln C$            &  -         &  $-11\pm7$                   &  (-14)     & $-6.7$                     &  -         & $-6.7$               &  -            \\
  $L$                & \ms        &  $21^{+9}_{-5}$              &  (16)      & $36.5$                     &  -         & $36.5$                 &  -            \\
  $P_{\rm GP}$       & days       &  $11.8\pm0.4$                &  (11.8)    & $11.4$                     &  -         & $11.4$                 &  -            \\
  $K_{s, b}$         & \ms        &  $10^{+1}_{-2}$              &  (10)      & $10^{+1}_{-2}$             &  (9)       & $10^{+1}_{-3}$             &  (11)         \\
  $\phi_{0, b}$      & rad.       &  $0.8^{+0.4}_{-0.3}$         &  (0.9)     & $0.8\pm0.3$                &  (0.8)     & $0.8^{+0.5}_{-0.3}$        &  (0.9)        \\
  $P_b$              & days       &  $2650\pm50$                 &  (2670)    & $2650\pm50$                &  (2670)    & $2650\pm60$                &  (2670)       \\
  $\omega_b$         & rad.       &  $3\pm2$                     &  (2)       & $3\pm2$                    &  (1)       & $3\pm2$                    &  (6)          \\
  $e_b$              & -          &  $0.01^{+0.06}_{-0.01}$      &  (0.01)    & $0.01^{+0.05}_{-0.01}$     &  (0.01)    & $0.01^{+0.06}_{-0.01}$     &  (0.01)       \\
  $K_{s, c}$         & \ms        &  -                           & -          & -                          & -          & $0.1^{+0.7}_{-0.1}$        &  (1.6)        \\
  $\phi_{0, c}$      & rad.       &  -                           & -          & -                          & -          & $3\pm2$                    &  (4)          \\
  $P_c$              & days       &  -                           & -          & -                          & -          & $80^{+700}_{-80}$          &  (10)         \\
  $\omega_c$         & rad.       &  -                           & -          & -                          & -          & $3\pm2$                    &  (4)          \\
  $e_c$              & -          &  -                           & -          & -                          & -          & $0.03^{+0.27}_{-0.03}$     &  (0.01)       \\
\hline
$\ln {\mathcal Z}$           & - & $-3289.24 \pm 0.02$ & & $-3287.76\pm0.10$ & & $-3287.12 \pm 0.13$ & \\
$\ln {\mathcal L_{\rm MAP}}$ & - & $-3227.4$ & & $-3231.5$ & & $-3220.8$ & \\
\enddata
  \tablecomments{Results of our \texttt{celerite} GP analysis on the full RV data set. The three models correspond to: GP \& 1-planet, in which all GP hyperparameters were left as free; GP \& 1-Planet, where three GP hyperparameters were fixed to their MAP values from the light curve analysis; and GP \& 2-Planets, where again three GP hyperparameters were fixed. Columns contain the median of the marginalized distribution of each parameter of interest, and uncertainties correspond to $16^{\rm th}$ and $84^{\rm th}$ percentiles. Values in parentheses ``()'' denote the MAP values. The bottom rows contain the log-evidences returned by the nested sampler and the log-likelihood of the MAP vector. The value and uncertainty of each $\ln {\mathcal Z}$ represent the median and standard deviation of three separate sampler runs, respectively. Jitter and offset parameters for each time series are omitted from the table. Note for the 2-Planets model, a significant number of samples involve the sampler effectively swapping between planet b and planet c. Rather than enforcing a prior to maintain ordering between the planets, we simply filtered samples for the statistics reported above for this model. Orbital parameters correspond to samples with $P_c < 2000$ days.}
\label{tab:rvresults}
\end{deluxetable*}

\subsection{RV Analysis}

\begin{figure*} 
\includegraphics[width=\linewidth]{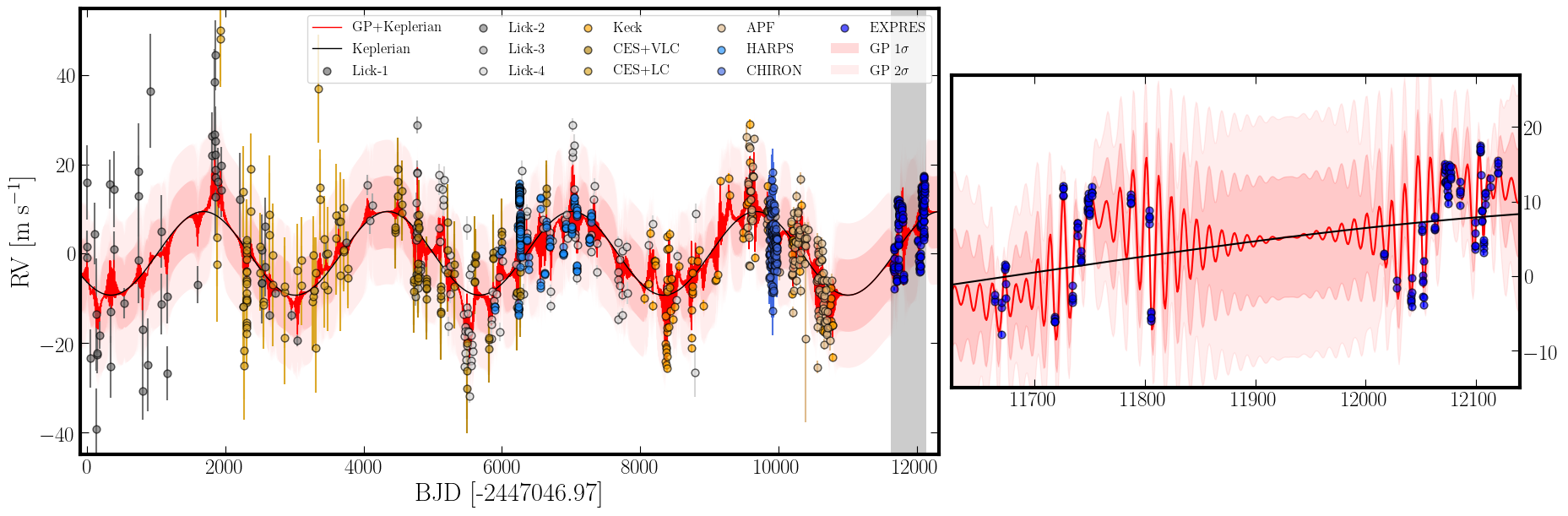}
\caption{Best-fit GP \& 1-Planet model that was fit to the combination of archival $\epsilon$ Eri RVs plus new EXPRES RVs (dark blue scatter points). The archival data set is identical to that analyzed by \citet{maw19}, with the addition of CHIRON RVs \citep{gig16}. For clarity, the MAP offset values ($\gamma_k$) have been subtracted from each RV data set. A zoomed-in panel of the EXPRES RVs is shown at right. The GP mean (red line) predominantly tracks the 7-year planetary signal (black line), with deviations owing to stellar activity. The quasi-periodic activity signal is more clearly seen in the zoomed-in panel. The period of oscillations is usually close to the periodic timescale hyperparameter, which was fixed to the 11.4 day stellar rotation period. By choosing a quasi-periodic kernel, the GP can accommodate small variations in period and amplitude, as well as gradual change in the activity signal's phase. The characteristic timescale for these variations is set by the parameter $L$. The GP $1\sigma$ and $2\sigma$ confidence intervals are depicted as lightly shaded regions around the GP mean. CES+LC and CES+VLC correspond to data obtained with the Coudé Echelle Spectrograph (CES) with the Long Camera (LC) and Very Long Camera (VLC), respectively. Lick data are color-coded by their corresponding upgrade epoch.}
\label{fig:allrvs}
\end{figure*}

We proceeded to model over $30$ years of radial velocities, with a focus on characterizing stellar activity, obtaining tight constraints on the orbit of the known planet $\epsilon$ Eri b, and searching for additional planets. \citet{maw19} showed that secular acceleration of $\epsilon$ Eri has negligible impact on derived parameters, and eventually excluded it from their model. Additionally, they partition the Hamilton/Lick RVs \citep{fis14} into four distinct data sets on account of instrument upgrades. We repeat these decisions in our analysis. In total, we jointly analyze 11 different RV data sets, which include distinct time series for each Lick epoch and the two cameras used in conjunction with Coudé Echelle (see Figure \ref{fig:allrvs}). Therefore, our entire RV model consists of the following: a GP activity component, which contributes one free hyperparameter $B$ corresponding to the covariance amplitude; a systemic velocity offset $\gamma_k$ and jitter term $s_k$ added in quadrature with each data point's uncertainty, for each RV data set $k\in \{1,2,3...11\}$; and five orbital elements $\{K_s, \phi_0, P, \omega, e\}$ for each Keplerian component, corresponding to semi-amplitude, phase of first epoch, orbital period, longitude of periastron, and eccentricity, respectively. The prior on $B$ was a log-uniform probability density function distribution ranging from 0.1 to 2500, or $\mathcal{LU}(0.1, 2500) \: (\ms)^2$. Priors on all $\gamma_k$ and $s_k$ were $\mathcal{U}(-30, 30) \: \ms$ and $\mathcal{LU}(0.01, 20) \: \ms$, respectively. The semi-amplitude $K_{s,b}$ prior was $\mathcal{LU}(0.1, 20) \: \ms$, and the orbital period $P_b$ prior was $\mathcal{LU}(2, 10000)$ days.

The fit results for parameters of interest are summarized in the second column of Table~\ref{tab:rvresults}. Our fitted orbital parameters for the 7-year planet agree with those of \citet{maw19} within $1\sigma$ uncertainties. The fitted RV curve is shown in Figure~\ref{fig:allrvs}, where $1\sigma$ and $2\sigma$ confidence regions are derived analytically from the GP posterior predictive distribution, with hyperparameters fixed at MAP values. We confirm that the orbit is consistent with circular \citep{maw19}. We fit an additional model of two planets with preconditioning of GP hyperparameters (Table~\ref{tab:rvresults}, third column). We fixed the GP hyperparameters $L$, $P_{\rm GP}$, and $\log C$ to photometry-derived values in order to reduce the fit dimensionality, and adopted the same priors on the second Keplerian parameters as used for the first Keplerian component. The Bayesian evidence prefers the two-planet model marginally, but insufficiently to justify the additional Keplerian parameters. The second planet's semi-amplitude is consistent with zero.

In Section~\ref{sec:specsims}, we more intensively study the activity-induced signal in the recent EXPRES RVs. The MAP orbital solution in the one-planet, preconditioned model (Table~\ref{tab:rvresults}, second column) serves as the RV baseline.

\section{Photospheric Brightness Variations} \label{sec:images}

Here, we model the spotted stellar surface of $\epsilon$ Eri with two different techniques: light curve inversion of TESS photometry and spot modeling with interferometry from the CHARA Array using the MIRC-X beam combiner.  Both techniques assume the star is spherical, as distortions are not expected for this star with $v \sin i = 2.93 \pm 0.5$ \kms \citep{gig16}.

\subsection{Light-curve Inversion Reconstructions} \label{sec:li}

While ground-based photometry (Section \ref{sec:APT}) is available for $\epsilon$ Eri, we focus here only on  the 2-minute cadence TESS light curve, which has significant overlap with both the EXPRES observations and our two MIRC-X observations.

\begin{figure*}[t!]
\begin{center}
\includegraphics[width=\linewidth]{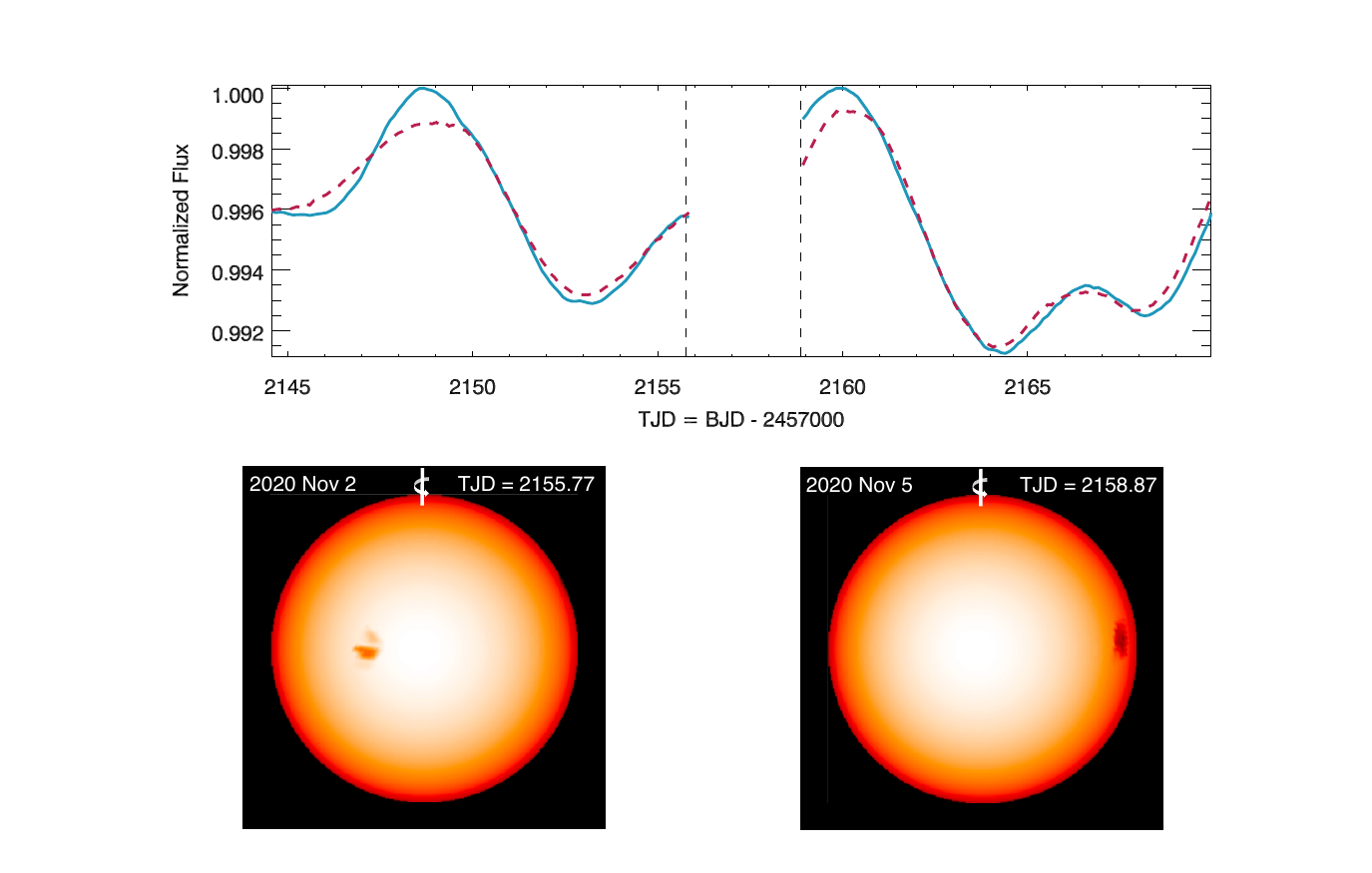}
\caption{Top:  Binned TESS light curve (solid blue line) and LI-reconstructed light curve (dashed red line) of $\epsilon$ Eri.  The binned light curve is normalized to the maximum value for each rotation.  Two stellar rotations were reconstructed using LI with the gap in TESS data separating the rotations. The dashed vertical black lines indicate the times of the interferometric MIRC-X observations discussed in the Section \ref{sec:interfmodels}. Bottom:  LI-reconstructed surfaces of the binned TESS light curve.  The visible pole is marked with a white line and the surface rotates counterclockwise around this pole. The surface temperature ranges from a spot temperature of $T_\mathrm{spot} = 4100$~K to a photospheric temperature of $T_\mathrm{phot} = 5100$~K.  The surfaces are shown at the times of the MIRC-X observations.}
\label{fig:LIpanel}
\end{center}
\end{figure*}

The Sector 31 TESS light curve encompasses just over two rotations of $\epsilon$ Eri based on our $P_\mathrm{rot} = 11.4$~days.  At the gap in data between the two orbits that make up Sector 31, we split the data into two rotation periods (2459144.5196172--2459155.8934423 and 2459158.8670810--2459169.9489975). The observations were averaged in 100 equally-sized bins in phase across each of the two rotation periods.   We reconstructed each rotation separately with the algorithm Light-curve Inversion \citep[LI;][]{har00}.  LI makes no \emph{a priori} assumptions of the starspot shape, number, or size and uses a modified Tikonhov regularizer to reconstruct the stellar surface \citep[for more details on LI see][]{har00,roe11,roe13}.   We provide the algorithm input parameters of the effective temperature $T_\mathrm{eff} = 5100$~K \citep{gig16}, starspot temperature $T_\mathrm{spot} = 4100$~K \citep[based on][]{ber05}, quadratic limb-darkening coefficients $a = 0.4258$ and $b = 0.1936$ \citep{cla18}, and  stellar inclination $i = 70^\circ$ \citep{gig16}, where equator-on viewing corresponds to $i = 90^\circ$.  We estimate the spot-to-photosphere brightness ratio  by integrating the radiation of blackbodies with the temperatures of the starspot and the photosphere over the spectral response function of the TESS bandpass \citep[as in][]{roe13}.  For $T_\mathrm{spot} = 4100$~K and $T_\mathrm{eff} = 5100$~K, the brightness ratio is $0.40$, i.e., the spot has $40\%$ of the brightness of the photosphere.   LI finds the optimum  reconstruction that simultaneously fits the prescribed spot-to-photosphere brightness ratio and the prescribed root-mean-square (rms) difference between the observed and model light curves.  We required the rms difference for these observations to be 0.0005 \citep[a value comparable to that used for applying LI to a \emph{Kepler} light curve;][]{roe16a}. Because of the small amplitude of the TESS light curve, this rms allows for the systematic deviations between the observed and reconstructed light curves shown in Figure \ref{fig:LIpanel}.  Reducing the rms value to improve the fit resulted in surfaces with features that are characteristic of overfitting \citep[e.g., elongated dark features that are not consistent with the structures analogous to starspots;][]{har00}.  The light curve of $\epsilon$~Eri  evolved on a timescale shorter than the stellar rotation period, so we emphasize that LI assumes that the light curve does not evolve in the light curve being inverted.  

\begin{figure*} 
\begin{center}
\includegraphics[scale=1.2]{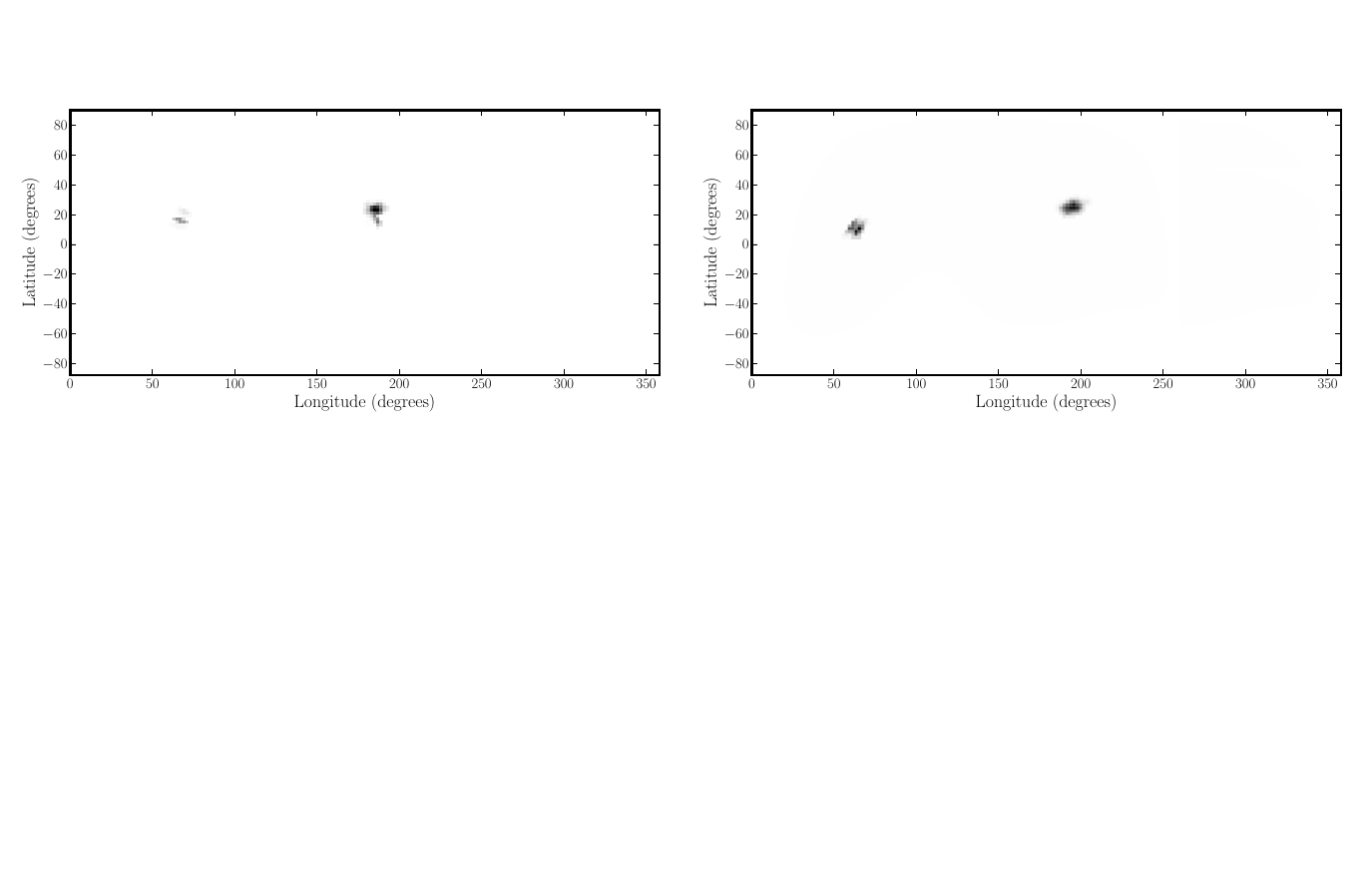}
 \vspace{-5.7cm}
\caption{Left:  LI-reconstructed pseudo-Mercator surface of $\epsilon$ Eri for the first rotation observed by TESS (ending before TJD 2459157).  At phase 0.0, as viewed by TESS, the center of the star is at longitude $90^\circ$. As time increases, the longitude of the center of the star decreases. Right:  LI-reconstructed pseudo-Mercator surface of $\epsilon$ Eri for the second rotation observed by TESS (starting after TJD 2459157). } 
\label{fig:TESS_psmerc}
\end{center}
\end{figure*}

In Figure \ref{fig:LIpanel}, we present the two LI-reconstructed surfaces from Sector 31 at the time of the MIRC-X observations.  Each of the two surface maps shown in Figure \ref{fig:LIpanel} features one prominent starspot.  The star rotates  $98^\circ$ from left to right between the two MIRC-X observations, so the spot seen on the left surface map is the same as the spot near the right limb in the right surface map.  We also present Figure \ref{fig:TESS_psmerc}, which shows the pseudo-Mercator projections of the entire stellar rotation from each portion of the TESS light curve.  In this presentation, JD 2459144.5196172 is considered phase 0.0, which corresponds to the a longitude of $90^\circ$ as center of the star, as seen by the observer.  As time increases, the longitude at the center of the star decreases because the star is assumed to rotate counterclockwise as seen from above the visible pole, while stellar longitude also increases in the counterclockwise direction. A second starspot is visible in both reconstructions.

\subsection{Interferometric Models of Starspots}
\label{sec:interfmodels}

For the two sets  of interferometric data, we individually fit a stellar surface to each one. The limb-darkening coefficient, the starspot location, and starspot size were fit separately to the data, as described below.  

For limb-darkening, we assume power-law limb darkening, which is defined as 

\begin{equation}
    I(\mu) = I_0 \mu^{\alpha},
\end{equation}
where $\mu$ is the cosine of the angle between the observer and the normal to the stellar surface, $I$ is the intensity, $I_0$ is the intensity at the center of the stellar surface, and $\alpha$ is the limb-darkening coefficient. 

For both sets of data, the starspot is defined to have a brightness of 61\% of that of the photosphere.  This value was estimated based upon the ratio of estimated spot and photospheric temperatures in $H$-band, the bandpass in which our MIRC-X observations were obtained.  The difference between the spot and photospheric temperatures is the same as was used with LI and was estimated based upon \citet{ber05}.

To construct a model of the star, we assumed the $H$-band, limb-darkened angular diameter measured by \citet{bai12}, $\theta_\mathrm{LD} =2.153$~mas.  While \citet{bai12} used a linear limb-darkening law \citep{mil21} and a model limb-darkening coefficient from \citet{cla95}, we used a power-law limb-darkening law, as described above, which was shown to be an appropriate model for interferometric observations by \citet{lac08} and described by \citet{hes97}.  To determine the best value of $\alpha$ for the limb-darkening, we fit a surface without spots to the data.  We selected $\alpha$ with the fit that had the lowest reduced $\chi^2$ value for the combination of the visibilities, closure phases, and triple amplitudes with the visibilities being weighted ten times more strongly than the other parameters to allow the closure phases and triple amplitudes to contribute to but not dominate the fit.  The visibility measurements were favored because the shape of the visibility curve more strongly constrains the limb-darkening parameter than the closure phases, which more strongly constrain surface asymmetries.  For the 2020 November 2 data, we found a best-fit limb-darkening coefficient of $\alpha = 0.27\pm0.02$.  For the 2020 November 5 data, we found $\alpha = 0.27\pm 0.01$.  Errors were determined with 100 bootstraps.  For example, for each bootstrap, from the 985 visibility data points for 2020 November 2, a point is randomly chosen 985 times with replacement, and this set, as well as similarly selected closure phases and triple amplitudes, is used to fit for $\alpha$.

\begin{figure*}
\begin{center}
\includegraphics[width=\linewidth]{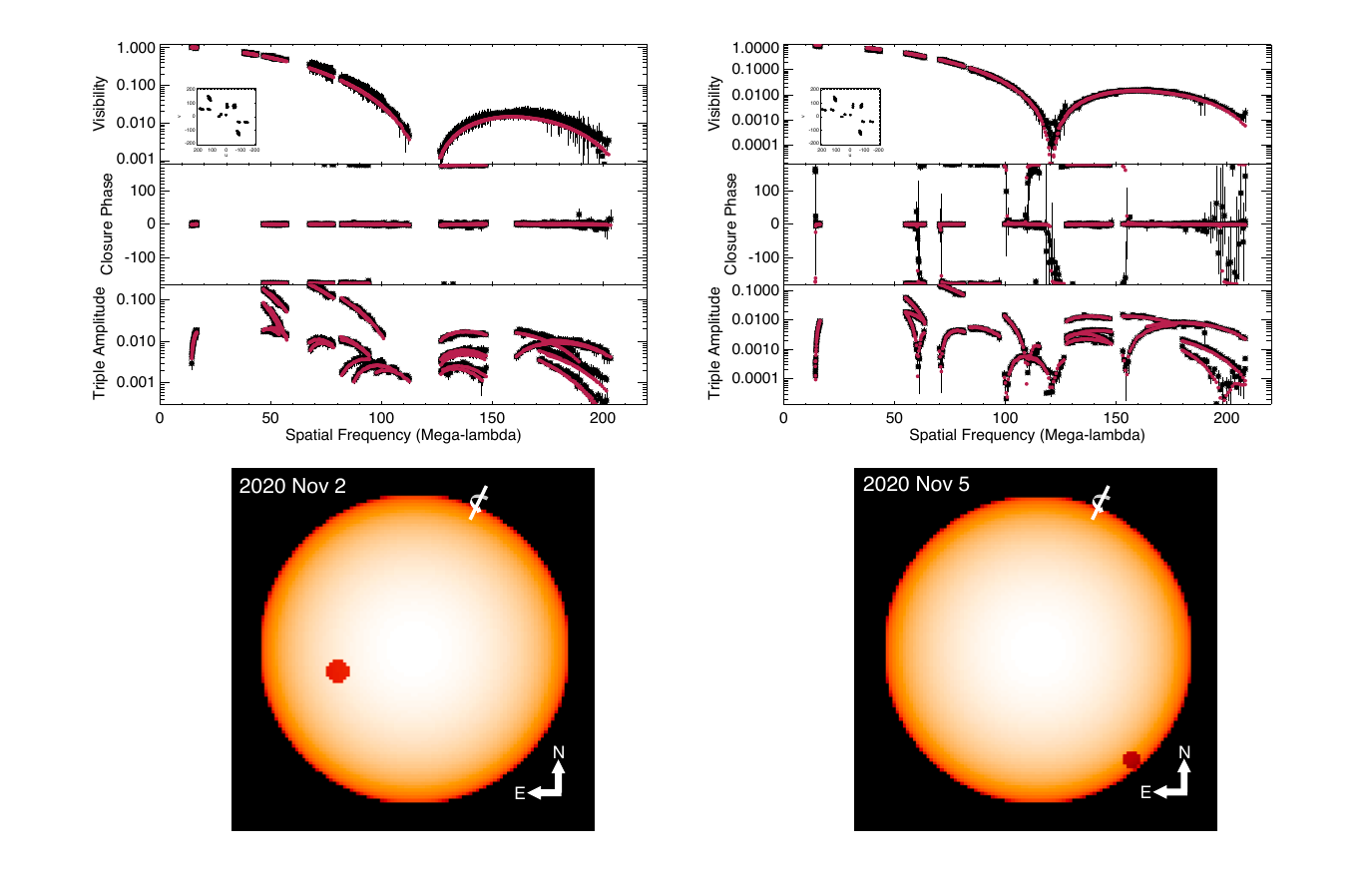}
\vspace{-.7cm}
\caption{Top:  MIRC-X observations of $\epsilon$ Eri from 2020 November 2 (left) and 2020 November 5 (right).  The observed data are shown in black with one baseline of each triangle shown in the closure phase and triple amplitude panels.  The model data for a limb-darkened, spotted surface are plotted in red. Bottom: The spotted, limb-darkened (power law) stellar surface models used to generate the interferometric data in red. The visible pole is marked with a white line and the surface rotates counterclockwise around this pole. As in Figure \ref{fig:LIpanel}, the surface temperature ranges from a spot temperature of $T_\mathrm{spot} = 4100$~K to a photospheric temperature of $T_\mathrm{phot} = 5100$~K.   The surface models are presented as they are on the plane of the sky where North (toward the celestial pole) is up  and East is to the left.   Because of the sparse $uv$ plane coverage, these starspot detections are marginal.  More data are needed for future studies aiming to image the star and determine the inclination and position angle of the rotational pole. }
\label{fig:mircxpanel}
\end{center}
\end{figure*}

We then assumed the stellar diameter and the limb-darkening and performed a grid search for the location of a simple, circular starspot with varying size to obtain the best fit for its location, which we show in Figure \ref{fig:mircxpanel}.  Ideally, to choose the best-fit spot location, we would select the location with the lowest reduced $\chi^2$ value for the combination of the visibilities, closure phases, and triple amplitudes with the closure phases being weighted ten times more strongly than the other parameters in this case because the closure phases are more sensitive to surface asymmetries than the other observables. Here, the weighting is chosen to allow the visibilities and triple amplitudes to contribute to but not dominate the fit.  While this best-fit method is valid for the night of 2020 November 2, when the starspot is near the center of the stellar disk and at nearly its largest contribution to the light curve, the starspot was on the limb on 2020 November 5 and at a much lower contribution to the light curve.  The best-fit spot location for 2020 November 5 is likely an artifact, but there is a local minimum that is consistent with the expected location of the spot based upon its previous location and the star having rotated through $98^\circ$ between the observations.  The local minimum spot location is shown in Figure \ref{fig:mircxpanel}.  The details of our recovery tests for simulated starspots using the $uv$ plane coverage of the 2020 November 5 data are in Appendix \ref{appendix:recoverytests}.  We note that the limited data available for these models makes these spot detections marginal, though consistent with the photometric observations. 

For the 2020 November 2 data, we performed 100 bootstraps, as described above, for the location of the starspot of the same size and darkness used for the best-fit data set to illustrate the quality of our fit.  The starspot locations are not easily quantified to error bars on the starspot location; therefore, we show the location of the starspots in Figure \ref{fig:starspoterrs} in Appendix \ref{appendix:recoverytests}. The locations of the starspots in these bootstraps are roughly consistent with the region surrounding the minimum of the $\chi^2$ surface of the best-fit data set (see Figure \ref{fig:chisq} also in Appendix \ref{appendix:recoverytests}.)

\subsection{Reconstruction and Model Differences} \label{sec:modeldiff}

While the TESS light curve and, consequently, the LI surface reconstructions of Figure \ref{fig:LIpanel} show more starspot evolution than is detectable in the interferometric models of Figure \ref{fig:mircxpanel}, we note that the TESS light curve is missing nearly all observations between the two interferometric observations. As a result, the reconstruction of the evolving spot that is visible at the time of both MIRC-X observations is more informed by the prior rotation (2020 November 2) and the following rotation (2020 November 5).  The LI surfaces give the impression that the morphology of the spot has changed in the three days between the MIRC-X observations.  While we assume some evolution has occurred, the differences between the spots as seen in Figure \ref{fig:TESS_psmerc} is potentially misleading and likely due to noise artifacts of the reconstructions.  Unfortunately, we are unable to resolve structure in the starspots with the interferometric observations for comparison.  

A notable difference between the LI reconstructions and the interferometric models are the locations of the spots.  LI has no ability to either constrain or determine the  position angle of the rotational pole in the plane of the sky.  Interferometric observations, however, do provide the opportunity to constrain the star's orientation on the sky, both the position angle and inclination given sufficient data.  Consistent with our LI reconstructions, we assume an inclination of $i = 70^\circ$ \citep{gig16} for the interferometric observations.  We investigated the position angle of the rotational pole of $\epsilon$ Eri by comparing the interferometric models to the surfaces reconstructed with LI.  The starspot longitudes taken from LI are reliable, but the starspot latitude may not be reliable due to lack of sufficient information from the input light curve(s).  Because we only used a single bandpass of data for these reconstructions, we assumed that the starspots are not necessarily located in latitude where they appear on the LI surfaces.  

\begin{figure*} 
\begin{center}
\includegraphics[trim=0 20 0 20,clip, scale=1.2]{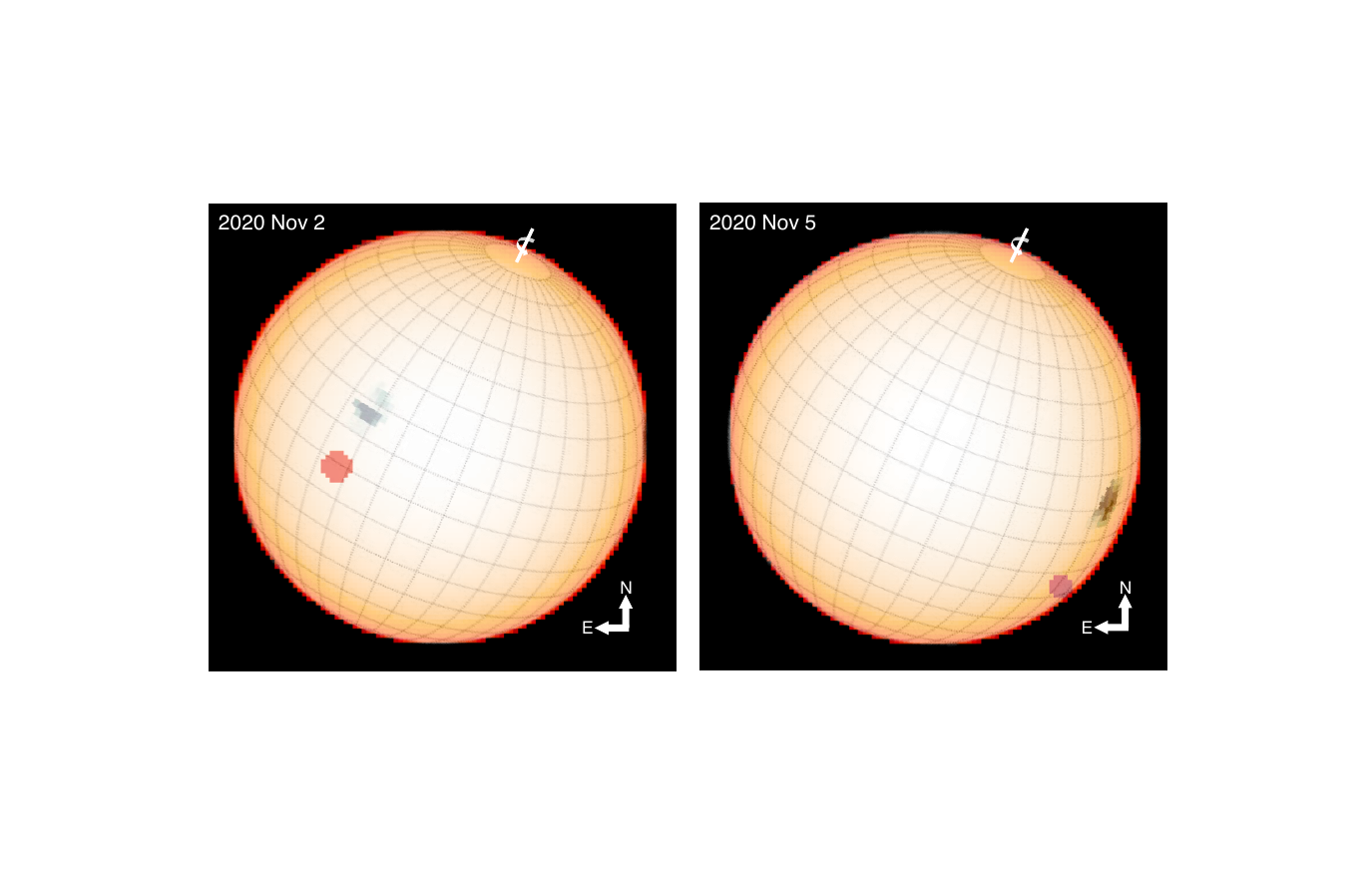}
 \vspace{-1.5cm}
\caption{Left: The best-fit interferometric fit for the 2020 November 2 data set with the LI reconstruction and its latitude and longitude grid overlaid, assuming $i = +70^\circ$.  The LI reconstruction is rotated $\sim335^\circ$ East of North in order to have the LI and MIRC-X starspots to lie on the same line of longitude.  The visible pole is marked with a white line and the surface rotates counterclockwise around this pole.    Right:  Similar to the left figure, but the 2020 November 5 data set.  This position angle determination is tentative due to the limited interferometric data. } 
\label{fig:MIRCXLIoverlay}
\end{center}
\end{figure*}

To approximate the position angle of $\epsilon$~Eri in the plane of the sky, we superimpose the interferometric models with the LI-reconstructed surfaces rotated in the plane of the sky.  For each snapshot interferometric data set, there are two orientations for which the modeled starspot and the corresponding LI-reconstructed starspot aligned in longitude; however only one orientation is consistent with both interferometric models.  Using the reconstructions and models for both 2020 November 2 and 5, we found that the data suggest that the position angle of the rotation axis of the star is $\sim335^\circ$, as presented in Figure \ref{fig:MIRCXLIoverlay}.  This assumes that the stellar inclination is $i=+70^\circ$, but the method used to determine the inclination in \citet{gig16} and LI both cannot make the distinction between $i = + 70^\circ$ and $i = -70^\circ$. Therefore, a position angle of $155^\circ$ is also possible if $i=-70^\circ$.  The position angle suggested in this analysis is tentative and must be confirmed with future data sets. Further details on this determination are found in Appendix \ref{appendix:recoverytests}.

\section{Spectroscopic Simulations} \label{sec:specsims}

Sector 31 TESS photometry span nearly 2.4 stellar rotations of $\epsilon$ Eri when it was also observed with EXPRES.  From these photometric observations, we reconstructed the stellar surface (Figures \ref{fig:TESS_psmerc}) and, here,  we discuss our method for identifying and isolating the RV signature of the starspots at any phase of rotation.

\subsection{Integrated Spectrum Over a Simulated Disk}

We simulated the stellar spectrum over time as the surface inhomogeneities reconstructed in the previous section rotate into and out of view for the time baseline of the TESS Sector 31 observations. This timeframe overlaps with 32 (of 164 total) spectra of $\epsilon$ Eri obtained with EXPRES between 2020 October and November. Our simulation is similar to the \texttt{SOAP} software \citep{bois2012}, which models perturbations to the spectral line profile induced by circular spots and plages, while accounting for limb-darkening, stellar rotation, and stellar geometry. Additional physics, including an improved limb-darkening and a model of convective blueshift suppression, were incorporated into \texttt{SOAP2} by \citet{dum14}. In a similar vein, 2D simulations of stellar disks have been used in studies of exoplanet transits \citep{cas2019}, with the purpose of investigating spectroscopic artifacts from center-to-limb variations and the Rossiter-McLaughlin effect. The premise is the same, in that the exoplanet occults a region of the star with its own local properties (i.e. white-light flux, Doppler shift). Our simulation code uses the physics and methodology used in \texttt{SOAP2}, but with EXPRES spectra for $\epsilon$ Eri instead of the solar spectrum.

A high-fidelity template spectrum of $\epsilon$ Eri was obtained by simultaneously fitting all 164 EXPRES spectra of the star with a B-spline regression similar to what is done by \texttt{SERVAL} \citep{zech18}. The B-spline is crucial for providing a smooth, continuous function to which we may apply arbitrarily small Doppler shifts. The disk model itself comprises an $80\times80$ pixel grid representing the visible surface of the star. Corner pixels beyond 1 $R_*$ remain empty throughout the following steps. Each pixel is assigned: (1) a finely-sampled ($R\sim800,000$) spectrum derived from the spectral template; (2) a flux weight (i.e., a relative contribution of flux to the integrated spectrum due to limb-darkening; the effect of spots is discussed below); and (3) a local velocity determined by stellar rotation. We approximate the spectra for various values of $\mu$ by assigning flux weights according to an appropriate limb-darkening law and use the appropriately Doppler-shifted EXPRES template spectrum. To generate a disk-integrated spectrum at each moment in time, we co-add the Doppler shifted and flux-weighted spectra from every pixel in the grid.

Critical to this study, the model is based on the LI surface reconstructions of the TESS light curve described in the previous section. First, the LI surface is interpolated from the existing spatial structure (approximately equal-area rectangular zones) onto a uniformly-spaced latitude/longitude grid (Figure~\ref{fig:TESS_psmerc}). We found 90 latitude divisions and 180 longitude divisions were sufficient for resolving the finest details in the LI surface. Rotation of the star is simulated by transforming the latitude and longitude coordinates, taking into account stellar inclination. Finally, the surface is mapped from 3D spherical coordinates to a 2D projection (via interpolation) on the pixel grid described above. At a given timestep, we multiply the flux weights of each pixel (determined by a quadratic limb-darkening law) by the relative brightness of the projected stellar surface, and then integrate the spectrum. In this way, we account for spots down-weighting local contributions to the integrated spectrum. As a consistency check, we summed the white-light flux of the projected disk at 30 equally-spaced intervals during a single rotation and recovered the relative brightness variations in the TESS light curve. The effects of resolution were explored, and we found the results did not change appreciably when more pixels were used to model the disk. RVs are obtained from the integrated spectrum via cross-correlation with the original stellar template.

\begin{figure*} 
\includegraphics[width=\linewidth]{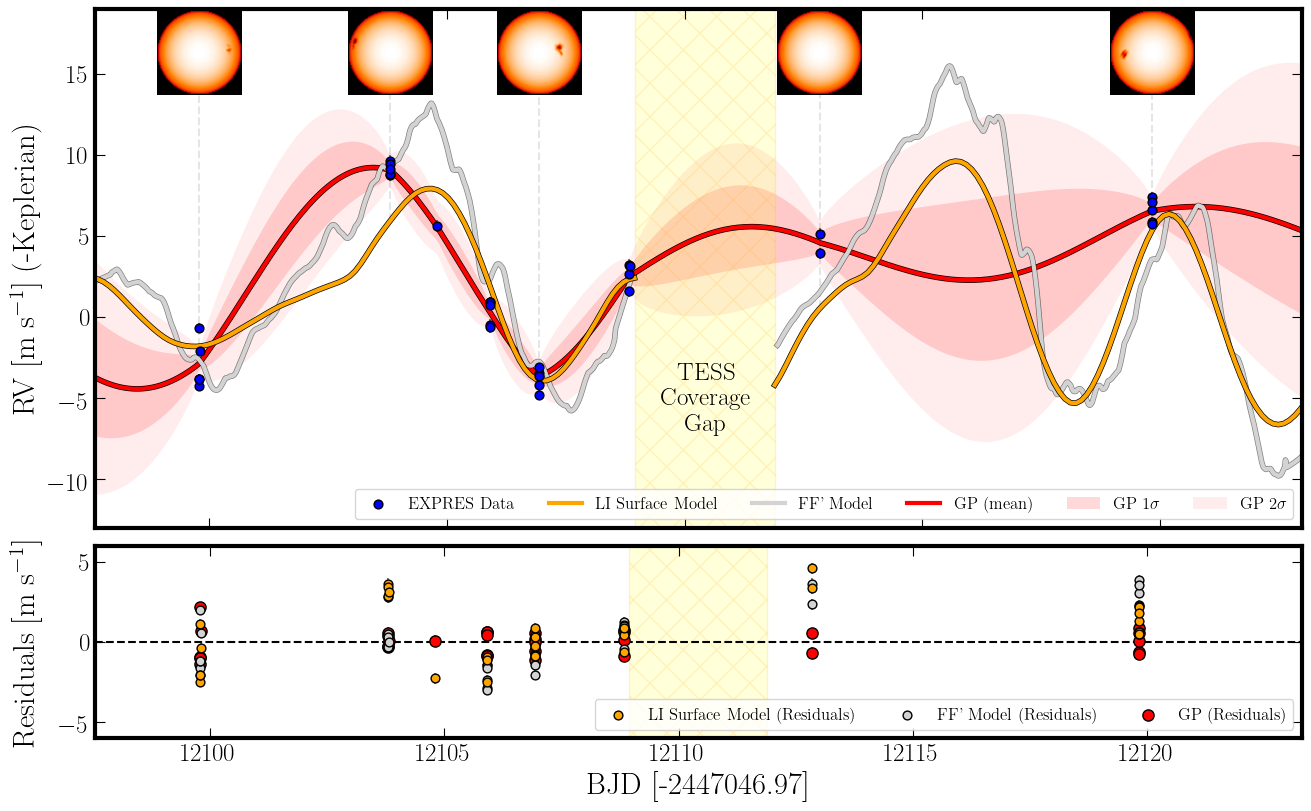}
\caption{Comparison of three models of activity-based RV variations. The red curve and shaded regions depict the GP model and confidence intervals (also shown in Figure~\ref{fig:allrvs}). The orange curve depicts the RV variations obtained by rotating the LI reconstructed stellar surface and integrating the stellar spectrum over a pixelated disk. For reference, snapshots of the disk model are shown for several selected timestamps of RV exposures. The features on the surface are responsible for the RV variations in the model, and a faint dashed line connects each snapshot to the corresponding point on the RV model curve. The gray curve represents an $FF'$ model \citep{aig12}, which serves as a benchmark for the LI surface model. The bottom panel shows residuals for the LI surface model, GP, and $FF'$ model in orange, red, and gray, respectively. The rms scatter of residuals after subtracting the GP is 0.76 \ms, which is largely a reflection of the EXPRES measurement uncertainties and intra-night scatter. The rms scatter of residuals after subtracting the LI surface model and best-fit offset is 1.98 \ms, compared to 2.20 \ms\ rms from subtracting the $FF'$ model. All three are a reduction from the original rms scatter of 4.72 \ms\ of this subset of RVs, which were selected based on their overlap with TESS photometry. In both the plot and rms calculations, the marginal Keplerian contribution has been subtracted.}
\label{fig:models}
\end{figure*}

Our LI surface model for the TESS Sector 31 data is shown in Figure~\ref{fig:models} and compared to both the actual EXPRES data, the GP model described in Section~\ref{sec:gp}, and an $FF'$ model \citep{aig12} discussed in the next section. The Keplerian component of the known planet has been subtracted, but this represented only a marginal RV trend of $\sim18$ \cms\ over the $\sim 20$-day timespan of the RVs considered. The RVs from the LI model contain relative velocity variations that arise from perturbations to the spectral line profile caused by the simulated spot. The actual EXPRES RVs are also relative velocities; however, the relative velocity offset in the EXPRES data changes as new data are acquired and will only asymptotically approach a constant value after all of the signals (i.e., the known planet in a 7-year orbit and the photospheric contributions) have been well sampled over all phases. The temporal baseline of EXPRES RVs, which is slightly over one year is not sufficient to reach that constant offset, so we derive and remove the best-fit offset between the simulated and observed RVs.

The subset of EXPRES RVs taken during the TESS Sector 31 observations has an rms scatter of 4.72 \ms. If we adopt corrections from the GP model, this scatter is reduced to 0.76 \ms. However, the GP model, by definition, is conditioned on the data to which it is fit. Therefore, the residuals reflect the measurement uncertainties added in quadrature with the `jitter' term $s_k$, added to account for intra-night scatter. The intra-night $s_k$ can arise from $p$-mode oscillations, granulation, or underestimation of formal uncertainties. This illustrates one of our primary concerns: even when the hyperparameters are conditioned on photometric observations,  the GP's flexibility (which is based on its parametrization and likelihood) allows it to conform to the RV measurements very closely, typically within $<1$ \ms\ from individual measurements or the mean RV of a given night.

The LI surface model, however, is completely independent of the EXPRES RVs. Aside from the constant RV offset discussed above, the LI surface model does not represent a fit to the RV data, but is derived strictly from the TESS photometry and the stellar spectrum template. After subtracting the RVs derived with the LI model and interpolated to the time of the EXPRES RV measurements, the residual rms of the EXPRES RVs decreases from 4.72 \ms\ to 1.98 \ms. This is a 58\% reduction in the RV scatter for $\epsilon$ Eri.

\section{Discussion} \label{sec:disc}

\subsection{Efficacy of the LI Surface RV Model}

We have shown that the LI surface stellar activity model presented in Section~\ref{sec:specsims} accounts for a significant portion of scatter in the EXPRES RV measurements. The success of this model marks an important step in robustly separating Keplerian and activity-related components in an RV time series; however, it is important to address the remaining $\sim 2$ \ms\ rms scatter and potential avenues for improving the model. For example, \citet{dum14} account for additional physics in their spot models, including inhibition of convective blueshift in regions affected by spots. Qualitatively, the net effect of accounting for convective blueshift in an equatorial spot is to break the symmetry in its corresponding RV signal, and push the RV signal towards more positive velocities. We experimented with a simple implementation of the convective blueshift effect by assigning a constant velocity offset inside the active region  \citep[e.g.,][adopt 350 \ms]{dum14}, but this addition did not improve the model fit. It is worth noting that $\epsilon$ Eri is a K2 dwarf, and the interplay between convective cells and the magnetic fields in active regions may differ from that in the Sun. It might also help to more carefully model spectra at different values of $\mu$. For example, \citet{cas2019} generate synthetic spectra for regions extending from the disk center to the limb. The change in effective temperature may have a minor impact on the cross-correlation and RV inference and it would be useful to retrieve faculae and accurate sizes for high-latitude features in the surface reconstructions; however, this will require longer-baseline interferometric measurements than are presently available. The most prudent next step is to improve the cadence of the radial velocities to allow for more dynamic modeling of the surface and to extend the time baseline of observations for both the RVs and the space-based photometry.

It is also worthwhile to explore additional physics. 
The full impact of stellar activity itself involves many components, including features on the rotating surface, as well as granulation, $p$-mode oscillations, and magnetic activity cycles \citep[][and references therein]{fis16}. While the amplitudes and timescales of different activity sources have been studied, the details needed to model their impact on RVs are not well-understood. 
Nevertheless, the growing body of literature adopting quasi-periodic GPs as an activity model supports the premise that, on timescales of several days, rotationally-modulated signals often have the highest amplitude effect on RVs, and are the most likely to generate false-positive Keplerian candidates. There is also strong precedent for inferring rotationally-modulated RV variations from photometry \citep{aig12, hay14}.

\subsection{Comparison to Gaussian Processes}

We modeled the full RV time series in Section~\ref{sec:gp} for the purpose of independently measuring orbital parameters of $\epsilon$ Eri b, evaluating the evidence for additional planets, and characterizing stellar rotation and typical spot lifetimes. Our model employed a GP via the \texttt{celerite} implementation. GPs are being applied as flexible stellar activity models by other groups as well in the current era of extreme precision spectroscopy \citep[e.g.][]{far20}. Our application of GPs closely resembles that of \citet{hay14}; however, one may simultaneously model activity indicators as done by \citet{raj15}. A number of even more advanced models are described in contributions to Zhao et al.\ (2021, in preparation), and a new GP model for inferring parameters governing the distribution of starspots, with potential applications to RV data sets, is presented by \citet{lug21}.

A particular advantage of the LI surface model is its derivation from an independent, photometric data set, which eliminates the possibility of Keplerian signals being absorbed into the model. We previously showed that  photometry-preconditioned GPs, however, are prone to conform to low-amplitude variations from a planet if that Keplerian signal is rejected under the Bayesian evidence comparison \citep[i.e., adding five additional parameters to the model does not sufficiently improve the log-likelihood of the model;][]{cab21}. This effect is especially true for sparse RV data sets. Another advantage is that the LI surface model has a direct, clearly interpretable correspondence with a physical characteristic of the star (i.e., resolved surface features), while GPs typically do not. One may use specific, physically-motivated GP kernels (e.g., quasi-periodic) and find repeating structure that corresponds to long-lived surface features, but it is a degenerate problem to invert the GP model and resolve surface features or other specific qualities of the star. In our case, the GP clearly does not correspond to the effects of surface features at all times and deviates at a $>2\sigma$ level from the LI surface model at most times (Figure~\ref{fig:models}). GP models that fit multiple, contemporaneous time series at once \citep[e.g.,][]{raj15, gil20} in implementations such as \texttt{pyaneti} \citep{bar2021} are not tested in this study, but may make for interesting future comparison against the LI model. Compared to the photometry-preconditioned GP used here, they may have reduced flexibility.

At present, GPs remain a useful model for stellar activity, which can greatly assist with identifying and constraining Keplerian signals.  However, accurately modeling RV variations with alternative and complementary data sets is an important goal, both for robustness against inadvertently removing Keplerian signals and for the interpretability of the activity model.

\subsection{Comparison to FF'}

The $FF'$ technique \citep{aig12} models RV variations based on contemporaneous photometry and provides a useful benchmark for our LI surface model. $FF'$ exploits the geometry of a spot moving across the stellar surface to remove explicit dependence on the rotation period, stellar inclination, and spot latitude. However, it is accurate only to first order in the presence of multiple spots, and neglects limb-darkening and spot projection effects. Extensions of the $FF'$ technique include \citet{raj15} and \citet{gig16}, which are not explored here. The full $FF'$ model is
\begin{equation}\label{eqn:ffprime}
        \Delta RV = \frac{\dot{\Psi}(t)}{\Psi_0} \Big(1-\frac{{\Psi}(t)}{\Psi_0} \Big)\frac{R_*}{f} + \Big(1-\frac{{\Psi}(t)}{\Psi_0} \Big)^2 \frac{\delta V_c \kappa}{f}\,,
\end{equation}
or equivalently
\begin{equation}\label{eqn:ffprime2}
        \Delta RV = \Delta RV_{\rm rot} + \Delta RV_{\rm c}\,,
\end{equation}
by denoting the left term as the rotation-related component and the right term as the convection-related component. In the above equation, $\Psi(t)$ is the light curve, and the constants $\Psi_0$, $\delta V_c, f, \kappa, R_*$ represent the disk's flux if no spots are present, the convective blue-shift inhibition within a magnetized region, reduction in flux for a spot at the center of the disk, the ratio of areas of the magnetized region and spot surface, and the stellar radius, respectively. Two parameters may be estimated directly from the light curve: $\Psi_0 \approx \Phi_{\rm max} + \sigma_{\Psi}$, where $\sigma_{\Psi}$ is the standard deviation of the light curve and $\Phi_{\rm max}$ is the light curve maximum; and $f \approx (\Psi_0 - \Phi_{\rm min})/\Psi_0$, where $\Phi_{\rm min}$ is the light curve minimum \citep{aig12}. 

The binned TESS light curves (Figure~\ref{fig:LIpanel}) were interpolated onto an oversampled grid and subsequently smoothed with a Savitzky–Golay filter, which allowed us to compute a smooth time derivative of the light curve $\dot{\Psi}(t)$. The window length was approximately $6\%$ of an 11-day TESS observing window. We manually varied the window size, and found the final residual rms scatter changed up to $\sim 20$ \cms. The adopted window size yielded the lowest residual rms scatter. We optimized $\delta V_c \kappa$ and a global model offset to fit the model to the subset of EXPRES data in Figure~\ref{fig:models}. The convection term in Equation~\ref{eqn:ffprime2} did not improve the fit, so $\delta V_c \kappa$ was fixed to 0 in the best-fit model. The convection term was significantly smaller than the rotation term in a case-study of HD~189733 \citep{aig12}, a star with a similar rotation period as $\epsilon$ Eri; however, suppression of convective blueshift may become the dominant process for slower rotators \citep[e.g.][]{hay14}.

The residual rms for the best-fit $FF'$ model is 2.20 \ms. The $FF'$ model has considerably more fine-structure than the LI surface model (Figure~\ref{fig:models}), which is due to either noise in the TESS light curve or smaller surface features unresolved by the inversion. However the largest features in both models are similar, which is expected since coincidentally only one spot is visible at most times, and the spots are small compared to the stellar disk; although the two models deviate by a few \ms\ midway through the first rotation, which may be due to the non-circular morphology of the visible spot. In addition to moderately outperforming $FF'$ by $\sim 20$ \cms\ reduction in rms, the LI surface model is more easily interpreted since it reveals the correspondence between specific spots and their RV perturbations. The remaining scatter in both sets of residuals indicates that more accurate modeling is needed for certain spot distributions (e.g., high-latitude features, or an odd-numbered multipole component), or that other features (e.g., faculae) or physics (e.g., granulation) are responsible for a significant portion of the activity signal.

\subsection{Imaging Sun-like Stars}

While the LI reconstructions are valuable in revealing the stellar surface's RV contribution, the surfaces are affected by the degeneracies of the light-curve inversion method.  The longitude of the starspots is well-constrained by the light curve, however, the starspot latitude is not.  With this method, information on constraining stellar latitude comes from limb-darkening.  Starspots at different latitudes will impact the light curve differently at different wavelengths because the limb darkening is different in different bandpasses.  For our $\epsilon$~Eri inversions, we use a single bandpass light curve from TESS.  Improving the latitude constraints of starspots being reconstructed with LI will require simultaneous light curves in a range of photometric filters.  

As stated in Section \ref{sec:intro}, reconstructing a stellar surface with Doppler imaging will provide more latitude information.  Although the method is being utilized for searching for hot Jupiters orbiting young stars \citep{hei21}, Doppler imaging is not an appropriate imaging method for main sequence stars, like $\epsilon$ Eri, because the stars' slow rotation does not provide sufficient spatial resolution.  

To date, interferometric aperture synthesis imaging, which unambiguously provides the latitude information, has not been performed on main-sequence stars.  However, our models of two data sets from the CHARA Array with the MIRC-X beam combiner indicate that detailed images of bright, spatially large main-sequence stars are possible with sufficient $uv$ plane coverage and prominent spots.  

Between the two interferometric observations, $\epsilon$ Eri rotated approximately a quarter rotation with the starspot present on both nights of interferometric data being the same spot.  Our models potentially constrain stellar orientation.  The interferometric data suggests that the star's rotation axis is oriented along the position angle approximately $\sim335^\circ$ East of North.  This was determined by rotating the LI reconstruction such that the MIRC-X observations would have the same longitude (see Section \ref{sec:modeldiff} and Appendix \ref{appendix:recoverytests}).  As we have only a small number of interferometric observations, confirmation of this position angle requires further observations.  

The starspot was observed approximately $20^\circ$ of latitude away from the starspot in the LI reconstructions.  This is in line with the expectation that surface reconstructed with LI will not necessarily place the starspots at the appropriate stellar latitudes.  LI favors smaller spots, which the regularizer identifies as ``smoother''; in combination with only one bandpass with limited limb-darkening constraints, LI will reconstruct the surface with starspots at the sub-Earth latitude.  For the $70^\circ$ inclination of $\epsilon$ Eri, the sub-Earth latitude is $+20^\circ$.  

While the starspot latitude appears to be lower than what LI predicts, we do not use this information to inform the simulated spectra, as our interferometric model for 2020 November 5 is not independently determined, and is informed by the previous observation and the LI surface due to the low contrast between the starspot and the limb (for more details, see Appendix \ref{appendix:recoverytests}).  However, these observations serve as a proof-of-concept that starspots can be interferometrically detected on a main-sequence star. 

Because the position angle is not independently determined for both nights of observation, we do not definitively state the star's orientation on the sky.  We also assume an inclination of $70^\circ$ after \citet{gig16}. The debris disk around $\epsilon$ Eri has been detected in a number of different studies that report inclinations ranging from about $20^\circ-30^\circ$ \citep{hol17,boo17,gre14,cha16,mac15}.  All of those studies give position angles around of $\sim0^\circ$, except for \citet{hol17}, who give $61^\circ\pm3^\circ$ East of North.  
The inclination of the plane of the planet is consistently found to be  $\sim30^\circ$ \citep{hat00,ben06,ref11}.  The longitude of the ascending node of the orbit was found to be $254^\circ\pm7^\circ$ by \citet{ben06} and $282^\circ\pm20^\circ$ by \citet{ref11}, both of which are in disagreement with the debris disk position angle measurements, which are defined by the major axis.  To compare whether the stellar equatorial plane is aligned with the debris disk and/or the planetary orbit, more MIRC-X observations are required to interferometrically measure both the star's position angle and inclination, from which an improved understanding of the system's orientation and evolution could be derived in addition to a better stellar image for our spectroscopic analysis.

\subsection{Conclusions and Future Work}

We have shown that if available, a light-curve inversion image of the stellar surface can provide crucial information for disentangling the signature of stellar activity from that of planets.  Obtaining a more accurate image will result in more accurate simulated RVs. 

We recommend both the RV and imaging observations be of high cadence.  \citet{cab21} emphasized the importance of high-cadence RV observations for modeling stellar activity with GPs and improving the RV precision for detecting planets.  Obtaining high-cadence observations for the complementary imaging is also vital in order to have an accurate image of the stellar surface from which RVs will be modeled.  Here, we obtained two short observations of $\epsilon$~Eri with the MIRC-X beam combiner.  More observations throughout a night would provide denser $uv$ plane coverage, and observations across the stellar rotation would allow for the entire stellar surface to be imaged in a method analogous to Doppler imaging and light-curve inversion.  The method developed here will work best when the imaging and RV observations are obtained contemporaneously to ensure that the same stellar surface evolution is being observed.  As seen for $\epsilon$~Eri in Figure \ref{fig:LIpanel}, the surface structures of main-sequence stars can change on timescales shorter than the stellar rotation.  

While each high-cadence data set investigating the stellar activity and potential for planetary companions is valuable, combining the data sets as we describe here allows for the strengths of EPRV and stellar surface imaging to be harnessed in a way otherwise only accessible for the Sun.  The unique value of solar studies is prior detailed knowledge of the surface features coupled with EPRV measurements.  We are working towards extending this to others stars by using high-cadence photometry to reconstruct the stellar surface and verify that the current interferometric capabilities can detect starspots on a main-sequence star.  Interferometric stellar images will provide unambiguous prior information about surface features to better understand the impact of photospheric activity on RVs.

\section*{ACKNOWLEDGEMENTS}

We are grateful to A.\ E.\ Szymkowiak for his role in the development of EXPRES; J.\ Ennis and A.\ Labdon for their role in the development of MIRC-X; and P.\ Gatkine, C.\ Trujillo, L.\ H.\ Wasserman, and M.\ West for contributing observations.  We thank the anonymous referee for thoughtful comments and insights that improved this paper.
These results made use of the Lowell Discovery Telescope at Lowell Observatory. Lowell is a private, non-profit institution dedicated to astrophysical research and public appreciation of astronomy and operates the LDT in partnership with Boston University, the University of Maryland, the University of Toledo, Northern Arizona University and Yale University.  Lowell Observatory sits at the base of mountains sacred to tribes throughout the region. We honor their past, present, and future generations, who have lived here for millennia and will forever call this place home.
Support for the design and construction of EXPRES is supported by the National Science Foundation (NSF) MRI-1429365, NSF ATI-1509436 and Yale University. We gratefully acknowledge support to carry out this research from NSF 2009528, NSF 1616086, NASA 17-XRP17 2-0064, the Heising-Simons Foundation, and an anonymous donor in the Yale alumni community.   
The CHARA Array is supported by the NSF under Grant No. AST-1636624 and AST-1715788. Institutional support has been provided from the GSU College of Arts and Sciences and the GSU Office of the Vice President for Research and Economic Development. MIRC-X received funding from the European Research Council (ERC) under the European Union’s Horizon 2020 research and innovation programme (Grant No. 639889). This research has made use of the Jean-Marie Mariotti Center \texttt{Aspro} service\footnote{Available at http://www.jmmc.fr/aspro}.
The APT photometric data were supported by NASA, NSF, Tennessee State University, and the State of Tennessee through its Centers of Excellence program.  This paper includes data collected by the TESS mission. Funding for the TESS mission is provided by the NASA's Science Mission Directorate.  
RMR acknowledges support from the Yale Center for Astronomy \& Astrophysics (YCAA) Prize Postdoctoral Fellowship, the Heising-Simons 51 Pegasi b Postdoctoral Fellowship, and NASA EPRV 80NSSC21K1034.  JMB was supported in part by NASA XRP 80NSSC21K0571. SK acknowledges support from an European Research Council Starting Grant (Grant agreement 639889).  SGJ acknowledges a partial support from the NASA TESS GI grant 80NSSC21K0243.

\facilities{TESS, DCT, CHARA, Fairborn APTs}

\newpage
\appendix 

\section{Posterior Distribution of the GP fit to APT Photometry}
\label{appendix:posterior}

\begin{figure} 
\centering
\includegraphics[width=0.6\linewidth]{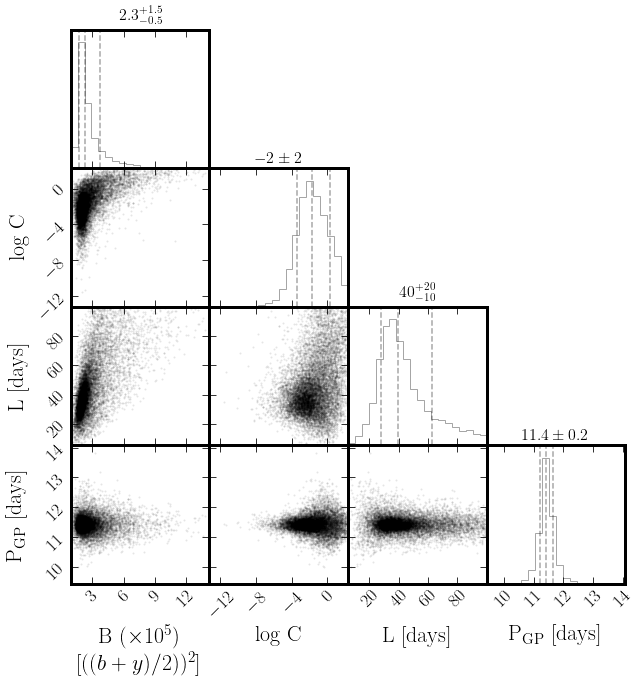}
\caption{Corner plot showing posterior distribution draws of GP hyperparameters (\texttt{celerite} quasi-periodic covariance kernel), after burn-in by the nested sampler. The $16\%, 50\%$, and $84\%$ quantiles are marked with dashed vertical lines in the marginalized histograms, which are also used to define the median values and uncertainties printed above each column.}
\label{fig:lccorner}
\end{figure}

Posterior draws of GP hyperparameters are shown in Figure~\ref{fig:lccorner}, based on the fit to APT photometry. Median values are listed with uncertainties corresponding to $16^{\rm th}$ and $84^{\rm th}$ percentiles. The MAP values of $P_{\rm GP}$, $L$, and $C$ were used in the pre-conditioned RV fits.

\section{Determining Starspot Location and Position Angle and Starspot Recovery Tests}
\label{appendix:recoverytests}

\begin{figure*}
\begin{center}
\vspace{-4cm}
\hspace{-3cm}
\includegraphics[scale=0.45]{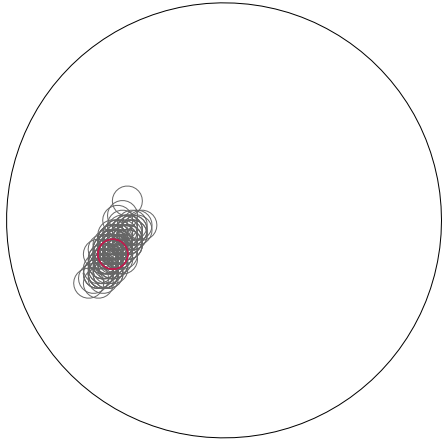}
\vspace{-2.cm}
\caption{Outline of the model stellar surface with the starspot location of each bootstrap indicated by a black circle.  Each circle is the size of the starspot used in the fit.  The thicker red circle is the location of the best-fit starspot location.  The stellar surface is oriented as in Figure \ref{fig:mircxpanel}}
\label{fig:starspoterrs}
\end{center}
\end{figure*}

\begin{figure} 
\begin{center}
\includegraphics[scale=1.2]{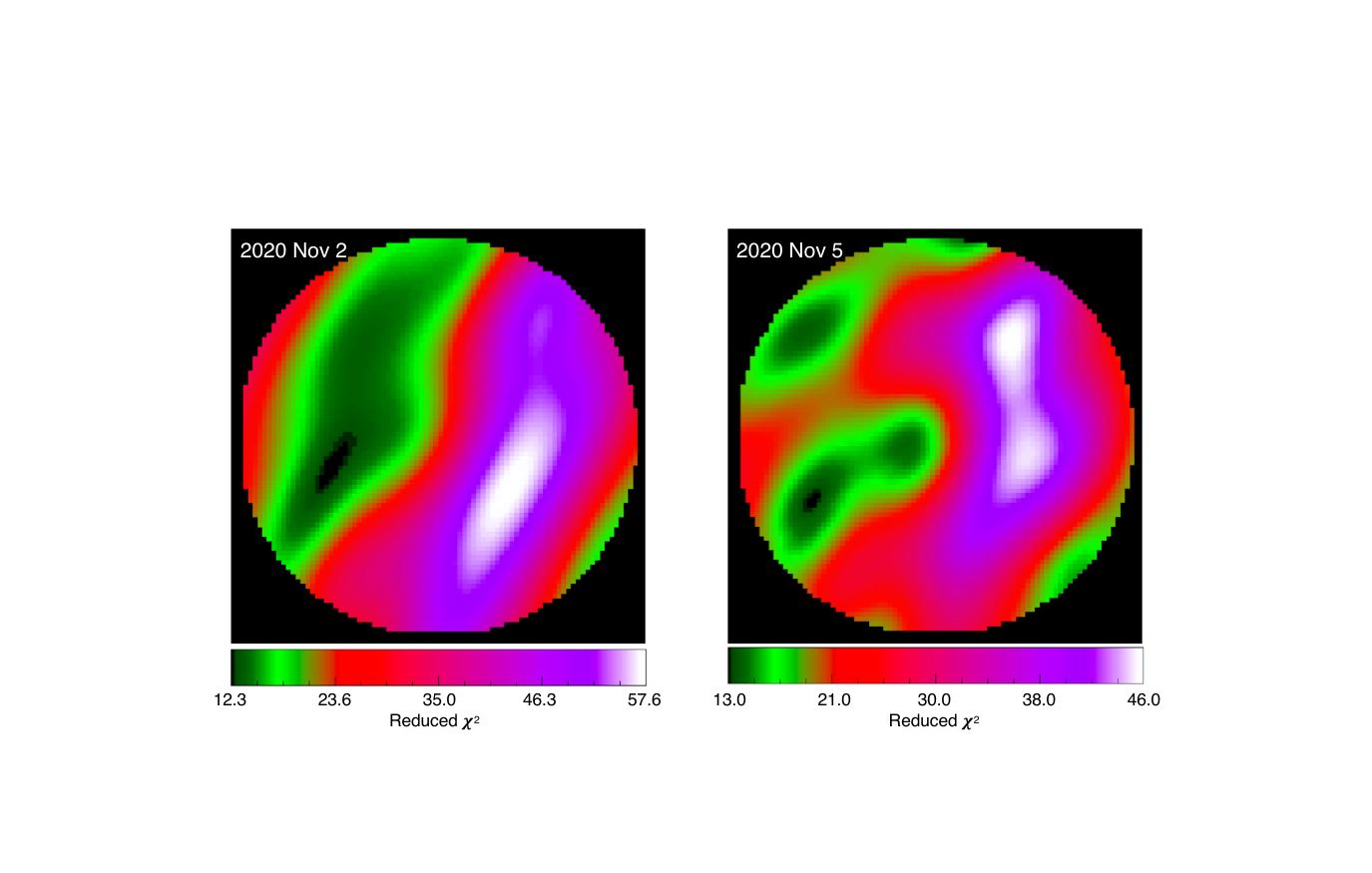}
 \vspace{-1.5cm}
\caption{Left: Reduced $\chi^2$ surface for detecting a starspot in the 2020 November 2 interferometric data. The color gradient was chosen to particularly emphasize the location of the global and local minima.   The minimum within the black region is the location of the model starspot shown in Figure \ref{fig:mircxpanel}. Right: Reduced $\chi^2$ surface for detecting a starspot in the 2020 November 5 interferometric data.  The black region is the global minimum, which our recovery tests show is likely to be an artifact, but the minimum on the lower right of the plot is aligned with our estimate of the starspot location given the position angles possible for the 2020 November 2 data and the LI reconstructions.  The spot at this location and its associated interferometric observations are shown in Figure \ref{fig:mircxpanel}.} 
\label{fig:chisq}
\end{center}
\end{figure}

In Section \ref{sec:interfmodels}, we discussed the interferometric models of $\epsilon$ Eri.  In order to show the robustness of the location of the starspot in the 2020 November 2 data, we performed 100 bootstraps where the visibility, closure phase, and triple amplitude data were chosen with replacement, as described in Section \ref{sec:interfmodels} for the limb-darkening parameter, $\alpha$.  We plot the results in Figure \ref{fig:starspoterrs} with each bootstrap being represented by a grey circle, of the approximate size of the starspot used in the fit and the best-fit spot location shown in red.  These bootstrap locations are approximately consistent with the region around the minimum of the $\chi^2$ surface shown in Figure \ref{fig:chisq}, described below.

As mentioned in Section \ref{sec:modeldiff}, we determined that the data suggest the position angle of $\epsilon$ Eri is $\sim335^\circ$ East of North.  This was found by superimposing the LI surfaces on top of the interferometric models and rotating the LI surfaces such that the LI starspot and the interferometric model starspot were aligned at the same longitude.  Because only one starspot was visible, there are two orientations that put the starspots at the same longitude.
For 2020 November 2, we found that the best-fit interferometric-modeled and LI-reconstructed starspots were aligned in longitude if the LI reconstruction was rotated $\sim75^\circ$ and $\sim335^\circ$ East of North in the plane of the sky.

However, the best-ft model for 2020 November 5 and the $~\sim98^\circ$ rotation of the star based upon the $P_\mathrm{rot} = 11.4$~days between the observations are not consistent.  In Figure \ref{fig:chisq}, we show the reduced $\chi^2$ surfaces for both 2020 November 2 and 5.  For 2020 November 2 there is a global minimum (represented by the black region within the reduced $\chi^2$ space), but for 2020 November 5, there are multiple local minima, and the lowest reduced $\chi^2$ value is again represented in black.  

Because we expect the starspot to be on the limb from the LI reconstructions, we investigated the ability of our interferometric spot-fitting model algorithm to detect starspots of varying size for a star with the angular diameter and limb darkening of $\epsilon$ Eri with the $uv$ plane coverage and telescope configuration of the 2020 November 5 observations.  

We aimed to recover a starspot in different locations and varying contributions to the overall brightness of the star (reductions in light from $0.1-6.4\%$).  We assigned the angular diameter to be $\theta_\mathrm{LD} = 2.153$~mas and limb-darkening coefficient $\alpha =0.27$, as above.  We used a circular starspot that had a spot-to-brightness ratio of $0.61$ and allowed its position and size to vary, but we required that the starspot had to stay within the star.  In Table \ref{tab:spottesttab}, we include the test and recovered spot parameters.

\begin{figure} 
\begin{center}
\includegraphics[angle=270]{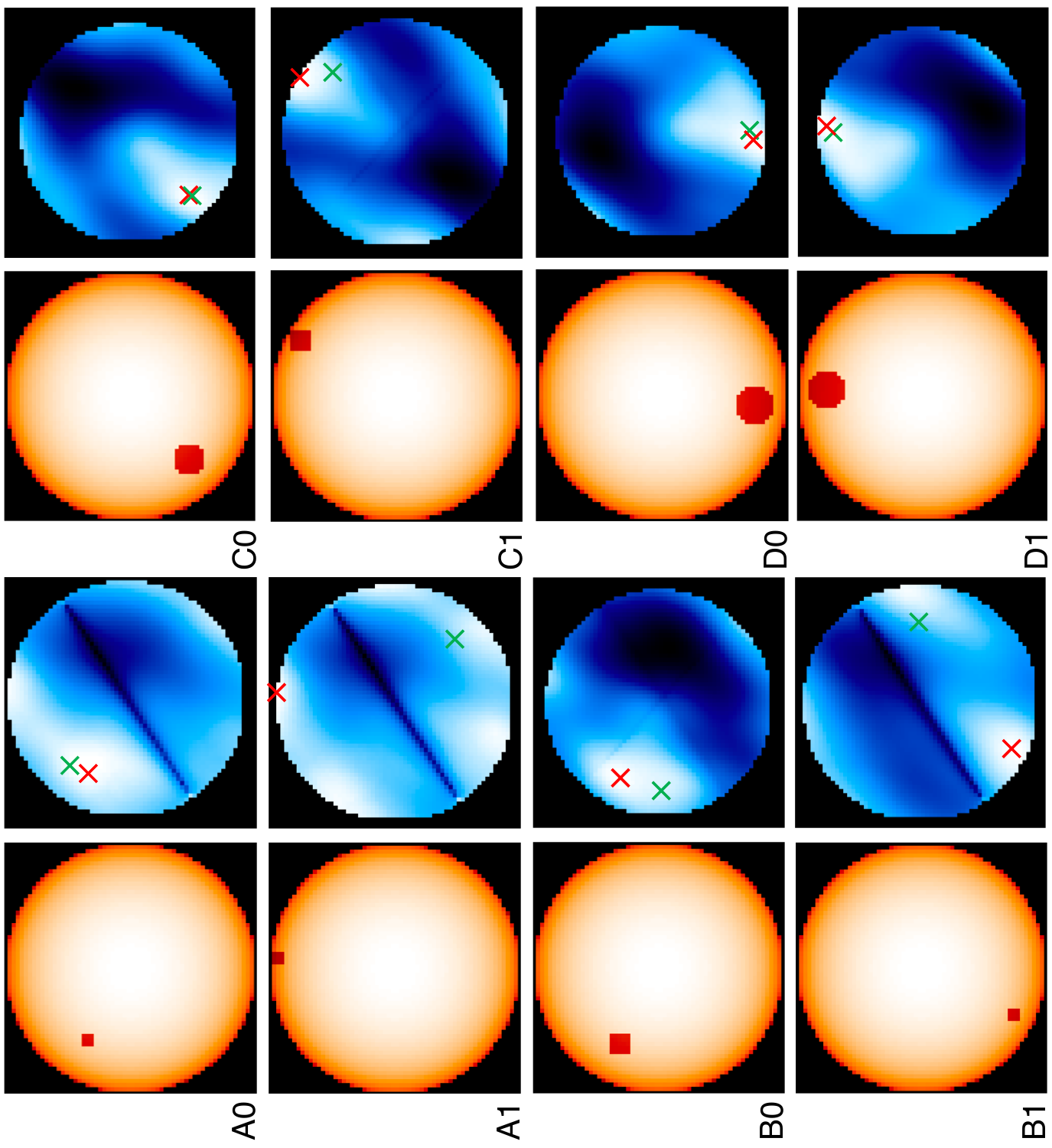}
\vspace{0.5cm}
\caption{First and third columns:  Best-fit model for the interferometric spot recovery test.  The surface temperature ranges from a spot temperature of $T_\mathrm{spot} = 4100$~K to a photospheric temperature of $T_\mathrm{phot} = 5100$~K.  Second and fourth columns: Reduced $\chi^2$ surface for the starspot size indicated by the best-fit model (lower values are white and higher values are dark blue).  The red $\times$ indicates where the center of the best-fit spot is located.  The green $\times$ indicates where the center of the spot was positioned.  Each row is labeled at the left and corresponds to the appropriately labeled row of test and recovered spot parameters in Table \ref{tab:spottesttab}.} 
\label{fig:spottests1}
\end{center}
\end{figure}

\begin{figure} 
\begin{center}
\includegraphics[angle=270]{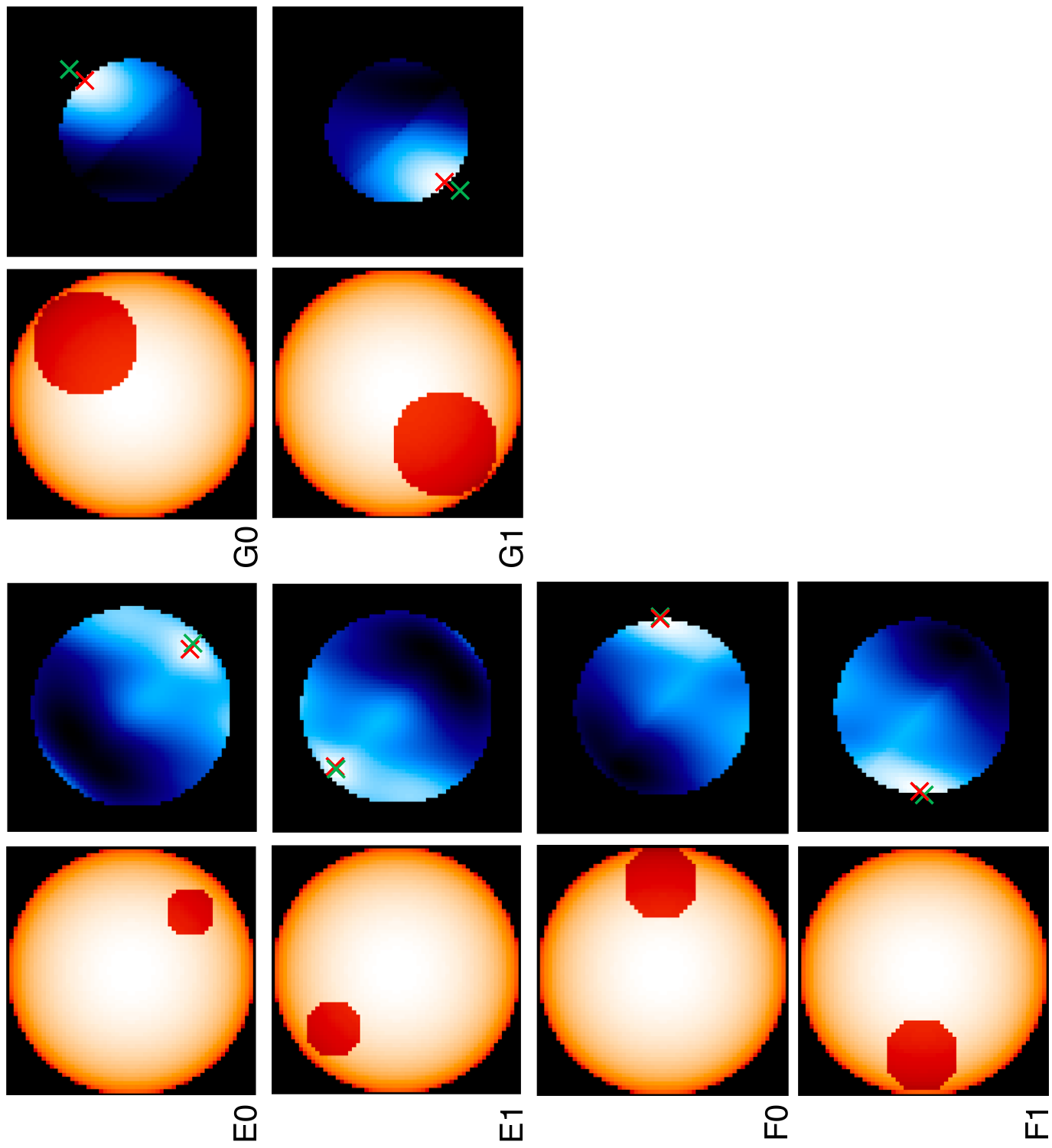}
 \vspace{0.5cm}
\caption{First and third columns:  Best-fit model for the interferometric spot recovery test.  Second and fourth columns:  Reduced $\chi^2$ surface for the starspot size indicated by the best-fit model (lower values are white and higher values are dark blue). The red $\times$ indicates where the center of the best-fit spot is located.  The green $\times$ indicates where the center of the spot was positioned.  Each row is labeled at the left and corresponds to the appropriately labeled row of test and recovered spot parameters in Table \ref{tab:spottesttab}.} 
\label{fig:spottests2}
\end{center}
\end{figure}

In Figures \ref{fig:spottests1} and \ref{fig:spottests2}, we show that we are able to reasonably accurately recover the location of the starspot when it  reduces the brightness of the star by at least $0.4\%$.  The reduced $\chi^2$ surfaces included in these figures are a combination between the visibilities, closure phases, and triple amplitudes with the closure phases being weighted ten times more than the other observables, as described above.  For spots that reduce the stellar brightness less than $0.4\%$, the best-fit solution is an artifact.  However, when examining the reduced $\chi^2$ space of the fit, the actual spot location is found to be a local minimum.  The starspot removing the smallest amount of light from the stellar surface explored here, $0.1\%$, is analogous to our estimations for the starspot on the limb.

Because the recovery tests indicate that the actual location of the starspot on 2020 November 5 may be recoverable as a local minimum, we considered these regions of the reduced $\chi^2$ space.  Extrapolating from the two possible position angles found for just 2020 November 2, $\sim75^\circ$  was rejected because the 2020 November 5 spot location was not in a local minimum, but the position angle of $\sim335^\circ$ placed the starspot in a local minimum, at the lower right of the right side of Figure \ref{fig:chisq}.  As a result, we include the starspot in this orientation and the associated model data in Figure \ref{fig:mircxpanel}.  More data are required, however, to confirm this orientation.  

Because we cannot verify the position angle and because the LI reconstructions of $\epsilon$~Eri indicate the presence of another starspot that is out of view on both 2020 November 2 and 5, we do not attempt to inform the photometric surfaces with the interferometric spot location.

\begin{deluxetable}{l c c c c c c}
\tabletypesize{\scriptsize}
\tablecaption{Interferometric Starspot Recovery Test Results}
\tablewidth{0pt}
\tablehead{
\colhead{Test Identifier} & \colhead{Assigned Distance } &  \colhead{Assigned Position } & \colhead{Assigned Brightness } & \colhead{Model Distance } &  \colhead{Model Position } & \colhead{Model Brightness } \\ \colhead{} & \colhead{from Center} &  \colhead{Angle ($^\circ$ E of N) } & \colhead{Decrease ($\%$)} & \colhead{from Center} &  \colhead{Angle ($^\circ$ E of N) } & \colhead{Decrease ($\%$)} 
}
\startdata
A0	&   0.75    &   45  &  0.10    &  0.74  &  61  &   0.13    \\
A1	&   0.75    &   225 &  0.10    &  1.01  &  356 &   0.10    \\
B0	&   0.75    &   90  &  0.20    &  0.75  &  65  &   0.35    \\
B1	&   0.75    &   270 &  0.20    &  0.93  &  152 &   0.12    \\
C0	&   0.75    &   135 &  0.40    &  0.79  &  133 &   0.62    \\
C1	&   0.75    &   315 &  0.40    &  0.95  &  331 &   0.30    \\
D0	&   0.75    &   180 &  0.80    &  0.83  &  173 &   0.93    \\
D1	&   0.75    &   0   &  0.80    &  0.83  &  358 &   0.91    \\
E0	&   0.75    &   225 &  1.60    &  0.74  &  223 &   1.50    \\
E1	&   0.75    &   45  &  1.60    &  0.76  &  45  &   1.98    \\
F0	&   0.75    &   270 &  3.20    &  0.75  &  270 &   3.33    \\
F1	&   0.75    &   90  &  3.20    &  0.75  &  90  &   3.39    \\
G0	&   0.75    &   315 &  6.40    &  0.58  &  313 &   7.19    \\
G1	&   0.75    &   135 &  6.40    &  0.63  &  133 &   7.19    \\
\enddata
\end{deluxetable}
 \label{tab:spottesttab}

\ 

\newpage
\bibliography{starspotpapers}

\end{document}